\documentclass[prd,email,preprint,showpacs,showkeys,preprintnumbers,amsmath,amssymb,nofootinbib]{revtex4}
\usepackage{graphicx}
\usepackage{amssymb}
\usepackage{epsfig}
\usepackage{dsfont}
\usepackage{amssymb,amsmath}
\usepackage{eufrak}
\usepackage{euscript}
\usepackage{latexsym}
\usepackage[latin1]{inputenc}
\usepackage{pstricks}

\def\csch{\mbox{csch}}
\newcommand{\be}{\begin{equation}}
\newcommand{\ee}{\end{equation}}
\newcommand{\ba}{\begin{eqnarray}}
\newcommand{\ea}{\end{eqnarray}}
\newcommand{\p}{\partial}

\skip\footins 20pt plus4pt minus4pt
\begin{document}

\title{Some aspects of quantum mechanics and field theory in a Lorentz invariant noncommutative space}

\author{Everton M. C. Abreu$^{a,b,c}$}
\email{evertonabreu@ufrrj.br}
\author{M. J. Neves$^a$}
\email{mariojr@ufrrj.br}

\affiliation{${}^{a}$Grupo de F\' isica Te\'orica e Matem\'atica F\' isica, Departamento de F\'{\i}sica,
Universidade Federal Rural do Rio de Janeiro\\
BR 465-07, 23890-971, Serop\'edica, Rio de Janeiro, Brazil\\
${}^{b}$LAFEX, Centro Brasileiro de Pesquisas F\' isicas (CBPF), Rua Xavier Sigaud 150,\\
Urca, 22290-180, RJ, Brazil\\
${}^{c}$Departamento de F\'{\i}sica, ICE, Universidade Federal de Juiz de Fora,\\
36036-330, Juiz de Fora, MG, Brazil\\
\bigskip
\today\\}
\pacs{11.15.-q; 11.10.Ef; 11.10.Nx}

\keywords{Noncommutativity; quantum mechanics; field theory}



\begin{abstract}


{\noindent We obtained the Feynman propagators for a noncommutative (NC) quantum mechanics
defined in the recently developed Doplicher-Fredenhagen-Roberts-Amorim (DFRA) NC background
that can be considered as an alternative framework for the NC spacetime of the early Universe.
The operators formalism was revisited and we applied its properties to
obtain a NC transition amplitude representation.
Two examples of DFRA's systems were discussed, namely, the NC free
particle and NC harmonic oscillator. The spectral representation of the propagator
gave us the NC wave function and energy spectrum.
We calculated the partition function of the NC harmonic oscillator
and the distribution function.
Besides, the extension to NC DFRA quantum field theory
is straightforward and we used it in a massive scalar field.
We had written the scalar action with self-interaction $\phi^{4}$ using the Weyl-Moyal product
to obtain the propagator and vertex of this model needed to perturbation theory.
It is important to emphasize from the outset is that
the formalism demonstrated here will not be constructed introducing a NC parameter in the system, as usual.
It will be generated naturally from an already NC space. In this extra dimensional NC space,
we presented also the idea of dimensional reduction to recover commutativity.} \\ \vspace{0.7cm}

\end{abstract}

\maketitle

\pagestyle{myheadings}
\markright{Quantum mechanics and field theory in a Lorentz invariant noncommutative space}

\section{Introduction}
\renewcommand{\theequation}{1.\arabic{equation}}
\setcounter{equation}{0}

There are theoretical evidences that make us to expect that at very small scales the spacetime acquires a foam-like
(fuzzy) structure \cite{hawking}.  This foam-like framework, at that time, was supposed to eliminate the problem
of infinities in quantum field theory.
However, it is well known now that noncommutativity,
due to its Planck scale feature, introduces this foam-like structure in spacetime,
which is characteristic of quantum gravity, where one should expect
a very large fluctuation of the metric and even of the topology of
the spacetime manifolds on short length scales \cite{hawking}.
Hence, to construct a NC spacetime implies directly to obtain this foam-like structure.

However, a long time before the foam-like space ideas of Hawking and Wheeler, the belief
that a noncommutative (NC) spacetime,
instead of a continuous one, could free quantum field theory
(QFT) from the divergences that habit within it.
This concept evolved through the years and
the first published work concerning a NC concept of spacetime was carried out in $1947$
by Snyder in his seminal paper \cite{snyder47}.
The need to control the ultraviolet divergences in quantum field theory (QFT) was the first motivation
to consider a NC spacetime. The main NC idea is that the spacetime coordinates $x^{\mu}\;(\mu=0,1,2,3)$
are promoted to operators $\hat{x}^{\mu}$ in order to satisfy the basic commutation relation
\begin{eqnarray} \label{xmuxnu}
\left[\,\hat{x}^{\mu}\,,\,\hat{x}^{\nu}\,\right]\,=\,i\,\ell\theta^{\mu\nu}\,\,,
\end{eqnarray}
where $\theta^{\mu\nu}$ is an antisymmetric constant matrix, and $\ell$ is a length scale.
The alternative would be to construct a discrete spacetime with a NC algebra.
Consequently, the coordinates operators are quantum observable that satisfy the uncertainty relation
\begin{eqnarray}\label{uncertainxmu}
\Delta \hat{x}^{\mu} \Delta \hat{x}^{\nu} \simeq \frac{\ell}{2} \,\theta^{\mu\nu} \; ,
\end{eqnarray}
it leads to the interpretation that noncommutativity of spacetime must emerge in
a fundamental length scale $\ell$, the {\it Planck scale}, for example.

However, Yang in \cite{yang47}, a little time later, demonstrated that Snyder's hopes in cutting off the infinities in QFT
were not obtained by noncommutativity. This fact doomed Snyder's NC theory to years of ostracism.
After the important result that the algebra obtained with a string theory embedded in a magnetic
background is NC, a new perspective concerning noncommutativity was rekindle \cite{seibergwitten99}.
Nowadays the NC quantum field theory (NCQFT) is one of the most investigated subjects about the description
of a physics at a fundamental length scale of quantum gravity \cite{QG}.


The most popular noncommutativity formalism consider $\theta^{\mu\nu}$
as a constant matrix, as we said before. Although it maintains the translational invariance,
the Lorentz symmetry is not preserved \cite{Szabo03}.
To heal this disease a recent approach was introduced
by Doplicher, Fredenhagen and Roberts (DFR) \cite{DFR}. It considers $\theta^{\mu\nu}$ as an ordinary
coordinate of the system in which the Lorentz symmetry is preserved.
Recently, it has emerged the idea \cite{Morita} of constructing an extension of the DFR spacetime
introducing the conjugate canonical momenta associated with
$\theta^{\mu\nu}$ \cite{Amorim1} (for a review the reader can see \cite{amo}).
This extended NC spacetime has ten dimensions:
four relative to Minkowski spacetime and six relative to $\theta$-space.
This new framework is characterized by a field theory constructed in a spacetime with extra-dimensions $(4+6)$,
and which does not need necessarily the presence of a length scale $\ell$ localized into the six dimensions of the $\theta$-space, where $\theta^{\mu\nu}$ has dimension of length-square. Besides the Lorentz invariance keeps maintained, obviously we hope that causality aspects in QFT in this $\left(x+\theta\right)$-spacetime must also be preserved \cite{EMCAbreuMJNeves2011}. The quantum field theory defined in DFR space is well known in the literature \cite{Carlson,Carone,Morita,Conroy2003,Saxell}, but it does not take into account any propagation of fields in the extra $\theta$-dimension. In this paper we are interested in analyzing the consequences of the propagation of fields in this $\theta$-direction.

By following this framework, a new version of NC quantum mechanics (NCQM) was introduced.  In this formalism
not only the coordinates $x^{\mu}$ and their canonical momenta $p^{\mu}$ are
considered as operators $\hat{x}^{\mu}$ and $\hat{p}^{\mu}$ in a Hilbert space ${\cal H}$, but the operators
of noncommutativity $\hat{\theta}^{\mu\nu}$ also have their canonical conjugate momenta operators
$\hat{\pi}^{\mu\nu}$ \cite{Amorim1,Amorim4,Amorim5,Amorim2}. All these operators belong to the same algebra
and have the same hierarchical level, introducing a minimal canonical
extension of DFR algebra, the so-called Doplicher-Fredenhagen-Roberts-Amorim (DFRA) formalism.
This enlargement of the standard set of Hilbert space operators permits us to consider the theory
to be invariant under the rotation group SO(D) treating it nonrelativisticaly.
Rotation invariance is underlying when we are treating
with nonrelativistic theories in order to depict consistently any physical system.
In \cite{Amorim4,Amorim5} the relativistic treatment is introduced,
which allows one to deal with the Poncar\`e invariance as a dynamical symmetry \cite{Iorio} in NCQM \cite{Amorim1}.

If the $\theta$ parameter is treated as a constant matrix, the non relativistic theory is not invariant
under rotation symmetry. This enlargement of the usual set of Hilbert space operators allows the theory
to be invariant under the rotation group $SO(D)$, as showed in detail in \cite{Amorim1,Amorim2}, where
the treatment is a non relativistic one. Rotation invariance in a non relativistic theory is the main
ingredient if one intends to describe any physical system in a consistent way.
In the first papers that treated the DFRA formalism, the main motivation, as we said before,
was to construct a NC standard QM.

The paper is organized as: the next section is dedicated (for self-containment of this work)
to a review of the basics of QM in this NC DFRA framework, namely, the DFRA algebra.
In the third section we apply the basics of DFRA NCQM to calculate the propagator via equation for the time
evolution of a quantum state.
In the fourth section we discussed the examples the free particle and the harmonic oscillator.  We calculated the partition function and the energy spectrum of the isotropic harmonic oscillator.
The fifth section is dedicated to field theory using the DFRA formalism in a well known standard model using the Weyl-Moyal product.
To finish, we discussed the results obtained and we made the final remarks and conclusions.




\section{Quantum mechanics in $DFRA$ NC space}
\renewcommand{\theequation}{2.\arabic{equation}}
\setcounter{equation}{0}

In this section, in order to maintain this work self-contained,  we will review the main steps published in \cite{Amorim1,Amorim4,Amorim5,Amorim2}.
Namely, we will revisit the basics of the NCQM defined in the DFRA space.
We assume that this space has $D \geq 2$ dimensions. The operators $\hat{x}_{i}$ $\left(i=1,2,...,D\right)$
and $\hat{p}_{i}$ are the position operator and its conjugated momentum, respectively. They satisfy the usual commutation relation
\begin{eqnarray}\label{xpcom}
\left[\hat{x}_{i},\hat{p}_{j}\right]=i\delta_{ij} \; ,
\end{eqnarray}
where we has adopted the Natural units $(\hbar=c=1)$ and $\ell=1$.
Having said that, in NCQM we can rewrite the commutation relation between the position operator, that is
\begin{eqnarray}\label{xxcom}
\left[\hat{x}_{i},\hat{x}_{j}\right]=i\theta_{ij} \; ,
\end{eqnarray}
where $\theta_{ij}$ is an antisymmetric matrix. In DFR formalism $\theta_{ij}$ is considered as a space
coordinate, and consequently it is promoted to a position operator $\hat{\theta}_{ij}$ in $\theta$-space.
Therefore this assumption leads us to a space with coordinates of position $\left(\hat{x}_{i},\hat{\theta}_{ij}\right)$,
in which $\theta_{ij}$ has $D(D-1)/2$ independent degrees of freedom. If $\theta_{ij}$ are coordinates,
the commutation relations are assumed to be
\begin{equation}\label{xthetacomm}
\left[\hat{x}_{i},\hat{x}_{j}\right]=i\hat{\theta}_{ij}
\hspace{0.4cm} , \hspace{0.4cm}
\left[\hat{x}_{i},\hat{\theta}_{jk}\right] = 0
\hspace{0.4cm} \mbox{and} \hspace{0.4cm}
\left[\hat{\theta}_{ij},\hat{\theta}_{k\ell}\right] = 0 \; .
\end{equation}
Moreover exist the canonical conjugate momenta operator $\hat{\pi}_{ij}$ associated with the operator $\hat{\theta}_{ij}$,
and they must satisfy the commutation relation
\begin{equation}\label{thetapicomm}
\left[\,\hat{\theta}^{ij},\hat{\pi}_{k\ell}\, \right] = i \delta^{ij}_{\,\,\,\,\,k\ell} \; ,
\end{equation}
where $\delta^{ij}_{\,\,\,\,\,k\ell}=\delta^{i}_{k}\delta^{j}_{\ell}-\delta^{i}_{\ell}\delta^{j}_{k}$.
In order to obtain consistency we can write that \cite{Amorim1}
\begin{equation}\label{ppicomm}
\left[\hat{p}_{i},\hat{p}_{j} \right] = 0
\hspace{0.2cm} , \hspace{0.2cm}
[\hat{p}_{i},\hat{\theta}_{jk}] = 0
\hspace{0.2cm} , \hspace{0.2cm}
[\hat{p}_{i},\hat{\pi}_{jk}] = 0 \; ,
\end{equation}
and this completes the DFRA algebra.

The Jacobi identity formed by the operators $\hat{x}_{i}$, $\hat{x}_{j}$
and $\hat{\pi}_{kl}$ leads to the nontrivial relation
\begin{equation}\label{nontrivialrel}
[[\hat{x}_{i},\hat{\pi}^{kl}],\hat{x}_{j}]- [[\hat{x}_{j},\hat{\pi}^{kl}],\hat{x}_{i}]
=- \delta^{ij}_{\;\;\;kl} \; ,
\end{equation}
which solution, not considering trivial terms, is given by
\begin{equation}\label{solutionnontrivial}
[\hat{x}_{i},\hat{\pi}^{jk}]=-{i\over 2}\delta_{il}^{\;\;\;jk}\hat{p}_{l} \,\,.
\end{equation}
It is possible to verify that the whole set of commutation relations listed above is indeed consistent
 under all possible Jacobi identities and the CCR algebras \cite{EMCAbreuMJNeves2011}. Expression (\ref{solutionnontrivial}) suggests that the shifted coordinate
operator \cite{Chaichian,Gamboa,Kokado,Kijanka,Calmet}
\begin{equation}\label{X}
\hat{X}_{i} = \hat{x}_{i}\,+\,{i\over 2}\hat{\theta}_{ij}\hat{p}^{j}\,\,,
\end{equation}
commutes with $\hat{\pi}_{kl}$.  The relation (\ref{X}) is also known as Bopp shift in the literature.
The commutation relation (\ref{solutionnontrivial})
also commutes with $\hat{\theta}_{kl}$ and $\hat{X}_{j} $, and satisfies a non trivial
commutation relation with $\hat{p}_{i}$ dependent objects, which could be derived from
\begin{equation}\label{Xpcomm}
[\hat{X}_{i},\hat{p}_{j}]=i\delta_{ij}
\hspace{0.6cm} , \hspace{0.6cm}
[\hat{X}_{i},\hat{X}_{j}]=0\,\,\quad \mbox{and} \quad [\hat{P}_i,\hat{P}_j]\,=\,0 ,
\end{equation}
where $\hat{P}_i=\hat{p}_i$ and the property $\hat{p}_{i} \hat{X}_{i}=\hat{p}_{i} \hat{x}_{i}$ is easily verified.
Hence, we see from these both equations that the shifted coordinated operator (\ref{X}) allows us to recover
the commutativity. The shifted coordinate operator $\hat{X}_{i}$ plays a fundamental role in NC
quantum mechanics defined in the $\left(x+\theta\right)$-space, since it is possible to form a basis with its eigenvalues.
This possibility is forbidden for the usual coordinate operator $\hat{x}_{i}$ since its components satisfy
nontrivial commutation relations among themselves (\ref{xpcom}). So,
differently from $\hat{x}_{i}$, we can say that $\hat{X}_{i}$ forms a basis in
Hilbert space. This fact will be important very soon.

With these definitions it seems interesting to study the generators of the group of rotations $SO(D)$.
It is a fact that the usual orbital angular momentum operator
\begin{eqnarray}\label{lij}
\hat{\ell}_{ij}\,=\,\hat{x}_{i}\hat{p}_{j}\,-\,\hat{x}_{j}\hat{p}_{i} \; ,
\end{eqnarray}
does not closes in an algebra due to (\ref{xxcom}), that is
\begin{equation}\label{llcomm}
\left[\hat{\ell}_{ij},\hat{\ell}_{kl}\right]=i\delta_{il}\hat{\ell}_{kj}-i\delta_{jl}\hat{\ell}_{ki}
-i\delta_{ik}\hat{\ell}_{lj}+i\delta_{jk}\hat{\ell}_{li}
+i\hat{\theta}_{il}\hat{p}_{k}\hat{p}_{j}-i\hat{\theta}_{jl}\hat{p}_{k}\hat{p}_{i}
-i\hat{\theta}_{ik}\hat{p}_{l} \hat{p}_{j}+i\hat{\theta}_{jk}\hat{p}_{l}\hat{p}_{i} \; ,
\end{equation}
and so their components cannot be $SO(D)$ generators in this extended Hilbert space.
It is easy to see that the operator
\begin{eqnarray}\label{LXij}
\hat{L}_{ij}\,=\,\hat{X}_{i}\hat{p}_{j}\,-\,\hat{X}_{j}\hat{p}_{i} \; ,
\end{eqnarray}
closes in the $SO(D)$ algebra. Besides, this result can be generalized to the
total angular momentum operator
\begin{eqnarray}\label{Jij}
\hat{J}_{ij}=\,\hat{X}_{i}\hat{p}_{j}\,-\,\hat{X}_{j}\hat{p}_{i}
+\hat{\theta}_{jl}\hat{\pi}^{l}_{\;i}-\hat{\theta}_{il}\hat{\pi}^{l}_{\;j} \; ,
\end{eqnarray}
that closes the algebra
\begin{eqnarray}\label{JijJklcomm}
\left[\hat{J}_{ij} ,\hat{J}_{kl} \right]=\delta_{il}\hat{J}_{kj}
-\delta_{jl}\hat{J}_{ki}-\delta_{ik}\hat{J}_{lj}+\delta_{jk}\hat{J}_{li} \; ,
\end{eqnarray}
and $\hat{J}_{ij}$ generates rotation in Hilbert space.

Now we return to the discussion about the basis in this NCQM.
It is possible to introduce a continuous basis for a general Hilbert space
searching by a maximal set of commutating operators. The physical coordinates
represented by the positions operators $\hat{x}_{i}$ do not commute and their eigenvalues
cannot be used to form a basis in the Hilbert space ${\cal H}_0$. This does not occur with
the shifted operators $\hat{X}_{i}$ (\ref{X}), and consequently, their eigenvalues are used
in the construction of such basis. Therefore one can use the shifted position operators
$\hat{X}_{i}$ as coordinate basis, although $\hat{x}_{i}$ be the physical position operator.
The noncommutativity of this space stays registered by the presence of the operator
$\theta$ as a spatial coordinate of the system.
A coordinate basis formed by the eigenvectors of $(\hat{X},\hat{\theta})$ can be introduced, and
for the momentum basis one chooses the eigenvectors of $(\hat{p},\hat{\pi})$.
Let $|X^{\prime},\theta^{\prime} \rangle=|X^{\prime}\rangle
\otimes |\theta^{\prime}\rangle$ and $|p^{\prime},k^{\prime} \rangle= | p^{\prime} \rangle \otimes |k^{\prime}\rangle$
be the position and momenta states in this $(x+\theta)$-space where the
fundamental relations involving each basis are
\begin{eqnarray}\label{eqautovaloresXtheta}
\hat{X}_{i}|X^{\prime},\theta^{\prime} \rangle
= X^{\prime}_{i}|X^{\prime},\theta^{\prime} \rangle
\hspace{0.5cm} , \hspace{0.5cm} \quad
\hat{\theta}_{ij}|X^{\prime},\theta^{\prime} \rangle
= \theta^{\prime}_{ij}|X^{\prime},\theta^{\prime} \rangle
\end{eqnarray}
\begin{eqnarray}\label{eqautovaloresppi}
\hat{p}_{i}|p^{\prime},k^{\prime} \rangle
= p^{\prime}_{i}|p^{\prime},k^{\prime} \rangle
\hspace{0.5cm} , \hspace{0.5cm} \quad
\hat{\pi}_{ij}|p^{\prime},k^{\prime} \rangle
= k^{\prime}_{ij}|p^{\prime},k^{\prime} \rangle \; ,
\end{eqnarray}
\begin{eqnarray}\label{completezarelXtheta2}
\int d^{D}{\bf X}^{\prime}d^{D(D-1)/2}\theta^{\prime} |{\bf X}^{\prime},\theta^{\prime} \rangle
\langle {\bf X}^{\prime},\theta^{\prime} | ={\bf 1} \; ,
\end{eqnarray}
\begin{eqnarray}\label{completezarelpkij}
\int \frac{d^{D}{\bf p}^{\prime}}{(2\pi)^{D}}
\frac{d^{D(D-1)/2}k^{\prime}}{(2\pi)^{D(D-1)/2}} |p^{\prime},k^{\prime} \rangle
\langle p^{\prime},k^{\prime} | ={\bf 1} \; ,
\end{eqnarray}
\begin{eqnarray}\label{ortognalityrelationXtheta}
\langle {\bf X}^{\prime},\theta^{\prime} |{\bf X}^{\prime\prime},\theta^{\prime\prime} \rangle =
\delta^{(D)}\!\left({\bf x}^{\prime}-{\bf x}^{\prime\prime}\right)\delta^{D(D-1)/2} \!
\left({\bf \theta}^{\prime}-{\bf \theta}^{\prime\prime}\right) \; ,
\end{eqnarray}
\begin{eqnarray}\label{ortognalityrelationpk}
\langle {\bf p}^{\prime},k^{\prime} |{\bf p}^{\prime\prime},k^{\prime\prime} \rangle = (2\pi)^{D}
\delta^{(D)}\!\left({\bf p}^{\prime}-{\bf p}^{\prime\prime}\right)
(2\pi)^{D(D-1)/2} \delta^{D(D-1)/2}\left(k_{ij}-k_{ij}^{\prime} \right)  ,
\end{eqnarray}
that are the eigenvalues equations, completeness relations and orthogonality, respectively.
Concerning the last relations and the operators representation we can write that \cite{Amorim1}
\begin{eqnarray}\label{XthetapXtheta}
\langle {\bf X}^{\prime},\theta^{\prime}|\hat{p}_{i}|{\bf X}^{\prime\prime},\theta^{\prime\prime} \rangle =
-i\frac{\partial}{\partial {\bf x}^{\prime i}} \delta^{(D)}\left({\bf x}^{\prime}-{\bf x}^{\prime \prime}\right)
\delta^{D(D-1)/2}\left({\bf \theta}^{\prime}-{\bf \theta}^{\prime \prime}\right) \; ,
\end{eqnarray}
and
\begin{eqnarray}\label{XthetapiXtheta}
\langle {\bf X}^{\prime},\theta^{\prime}|\hat{\pi}_{ij}|{\bf X}^{\prime\prime},\theta^{\prime\prime} \rangle =
\delta^{(D)}\!\left({\bf x}^{\prime}-{\bf x}^{\prime \prime}\right)
\; (-i)\frac{\partial}{\partial {\bf \theta}^{\prime ij}} \delta^{D(D-1)/2}\left({\bf \theta}^{\prime}
-{\bf \theta}^{\prime \prime}\right) \; .
\end{eqnarray}
It is important to pay attention to the notation. It is obvious that the prime and double prime notation indicates two different points in $(x+\theta)$-space.
But the meaning is the same, i.e., two different points in $(x+\theta)$-space.

The transformations (\ref{XthetapiXtheta}) from one basis to the other are constructed using extended Fourier transforms.
The wave plane defined in this $(x+\theta)$-space is obtained by internal product between position
and momentum states
\begin{eqnarray}\label{waveplane}
\langle {\bf X}^{\prime},\theta^{\prime} |{\bf p}^{\prime\prime},k^{\prime\prime} \rangle =
e^{i\left({\bf p}^{\prime\prime}\cdot {\bf x}^{\prime}+k^{\prime\prime}\cdot\theta^{\prime}
\right)} \; ,
\end{eqnarray}
where ${\bf p}^{\prime\prime}\cdot{\bf x}^{\prime}+k^{\prime\prime}\cdot\theta^{\prime}=
p_{i}^{\prime\prime}x^{i \; \prime}+k_{ij}^{\prime \prime}\theta^{ij \prime}/2$,
and we have used the property of scalar product ${\bf p}\cdot{\bf X}={\bf p}\cdot{\bf x}$.
Others properties of this NCQM are explored in more details
in \cite{Amorim1}.
\section{Feynman propagator in NC quantum mechanics}
\renewcommand{\theequation}{3.\arabic{equation}}
\setcounter{equation}{0}

To apply the basics of DFRA NCQM to a path integral formalism,
for simplicity we consider a space of $D=3$, {\it i.e.}, we have three independent coordinates
associated to $\theta_{ij}$ plus three usual position coordinates $\hat{x}_{i} \; (i=1,2,3)$.
The indices ($ij$) in $\theta_{ij}$ indicate one point in $\theta$-space formed by $(\theta_{12},\theta_{23},\theta_{31})$ coordinates. The same notation will be used for the
momentum $k_{ij}$, say $(k_{12},k_{23},k_{31})$. Thus we will work in a six dimensional space.

The time evolution of a quantum state $|\psi\rangle$ is governed
by the dynamical equation
\begin{eqnarray}\label{eqpsi}
i\frac{d}{dt}|\psi(t)\rangle=\hat{H}|\psi(t)\rangle \; ,
\end{eqnarray}
where $\hat{H}$ is the Hamiltonian operator. The solution of equation (\ref{eqpsi}) is the
expression
\begin{eqnarray}\label{solutionPsit}
|\psi(t)\rangle = e^{-i\hat{H}(t-t^{\prime})} |\psi(t^{\prime})\rangle  \ ,
\end{eqnarray}
that represents the time evolution of a state $|\psi\rangle$ in a time interval $t-t^{\prime}>0$ between two points $({\bf X},\theta)$ and $({\bf X}^{\prime},\theta^{\prime})$
in this $(x+\theta)-$space, and the Hamiltonian operator $\hat{H}$ is considered time independent. The coordinates       like $\theta$ and $\theta^{\prime}$ indicate that we are considering two different points in space, for instance, It is like the prime and double prime in Eqs. (2.19)-(2.23) and so on.

Let $|{\bf X},\theta\rangle$ be a position quantum state in $({\bf X},\theta)$, we operate it in
(\ref{solutionPsit}) to obtain
\begin{eqnarray}\label{psiXbThetab}
\langle {\bf X},\theta|\psi(t)\rangle=\langle {\bf X},\theta|e^{-i\hat{H}(t-t^{\prime})}|\psi(t^{\prime})\rangle \; ,
\end{eqnarray}
and introducing the identity (2.17)
%
%
we have
%
%
%
\begin{eqnarray}\label{psiXbthetabGpsiXathetaa2}
\psi({\bf X},\theta;t)=\!\!\int d^{3}{\bf X}^{\prime} d^{3}\theta^{\prime} \; K({\bf X},\theta;{\bf X}^{\prime},\theta^{\prime};t-t^{\prime})
\;
\psi({\bf X}^{\prime},\theta^{\prime};t^{\prime}) \; , \; \; \;
\end{eqnarray}
which provides the transition of the particle wave-function $\psi$ between the points $({\bf X},\theta)$ and $({\bf X}^{\prime},\theta^{\prime})$ in the $(x+\theta)$ space, and $K$
is the Feynman propagator of this transition
\begin{eqnarray}\label{KXaXb}
K({\bf X},\theta;{\bf X}^{\prime},\theta^{\prime};t-t^{\prime}):=\langle {\bf X},\theta|e^{-i\hat{H}(t-t^{\prime})}|{\bf X}^{\prime},\theta^{\prime}\rangle \; . \;\;\;
\end{eqnarray}
Hence, the propagator (\ref{KXaXb}) satisfies the Green equation
\begin{equation}\label{EqGreenG}
\left\{i\frac{\partial}{\partial \tau}-\hat{H}\left(\hat{X}_{i},\hat{p}_{i};\hat{\theta}_{ij},\hat{\pi}_{ij}\right)\right\}
K\left(X,\theta;X^{\prime},\theta^{\prime};\tau \right)=
\delta^{(3)}({\bf x}-{\bf x}^{\prime})\delta^{(3)}(\theta-\theta^{\prime})\delta(\tau) \; ,
\end{equation}
where we assumed the condition
\begin{eqnarray}
K\left(X,\theta;X^{\prime},\theta^{\prime};\tau \right)=0
\hspace{0.4cm} \mbox{when} \hspace{0.4cm}
\tau=t-t^{\prime}<0 \; .
\end{eqnarray}
The Hamiltonian operator $\hat{H}$ is a function of the position in the six-dimensional space
$(\hat{X}_{i},\hat{\theta}_{ij})$ and of the momenta operators $(\hat{p}_{i},\hat{\pi}_{ij})$
discussed in the last section. The spectral representation of the propagator is obtained
by inserting the complete set
\begin{eqnarray}
{\bf 1}=\sum_{n_{i},\widetilde{n}_{i}} | n_{i}; \widetilde{n}_{i} \rangle \langle n_{i}; \widetilde{n}_{i} |
= \sum_{n_{i}} | n_{i} \rangle \langle n_{i} | \sum_{\widetilde{n}_{i}} | \widetilde{n}_{i} \rangle \langle \widetilde{n}_{i} |
\end{eqnarray}
in the definition (\ref{KXaXb}), so we can write that
\begin{eqnarray}\label{Kesp}
K({\bf X},\theta;{\bf X}^{\prime},\theta^{\prime};\tau):=\Theta(\tau)
\sum_{n_{i}} \Phi_{n_{i},\widetilde{n}_{i}}({\bf X},\theta) \Phi_{n_{i},\widetilde{n}_{i}}^{\ast}({\bf X},\theta)
e^{-iE_{n_{i},\widetilde{n}_{i}}\tau} \; ,
\end{eqnarray}
where $\Phi_{n_{i},\widetilde{n}_{i}}({\bf X},\theta):=\phi_{n_{i}}({\bf X})\xi_{\widetilde{n}_{i}}(\theta)$, and
$\phi_{n_{i}}({\bf X}):=\langle {\bf X}| n_{i} \rangle$ ,
$\xi_{\widetilde{n}_{i}}(\theta):= \langle \theta |\widetilde{n}_{i} \rangle $ are the wave functions of the system, and
$E_{n_{i},\widetilde{n}_{i}}$ the energy spectrum. The eigenvectors
$|n_{i}; \widetilde{n}_{i} \rangle=|n_{i}\rangle \otimes |\widetilde{n}_{i}\rangle $ are the eigenstates of spectrum
energy, say $\hat{H}|n_{i}; \widetilde{n}_{i} \rangle=E_{n_{i},\widetilde{n}_{i}}|n_{i}; \widetilde{n}_{i} \rangle$.  It is important to notice that we are constructing these eigenvectors and its respective eigenstates in the $(X,\theta)$ NC space.  Namely, in the extended NC Hilbert space ${\cal H}'$ so that the general Hilbert space is given by ${\cal H}(X,\theta)\,=\,{\cal H}_0\,\oplus\, {\cal H}'$.

By using the definitions of operators of the DFRA algebra
and basis on the Hilbert space ${\cal H}$, the representation of the Feynman as a functional
integral over configuration space is
\begin{eqnarray}\label{intFuncDxDtheta}
\langle {\bf X},\theta|e^{-i\hat{H}(t-t^{\prime})}|{\bf X}^{\prime},\theta^{\prime}\rangle=
N\int \frac{{\cal D}{\bf X}}{(2\pi)^3} \frac{{\cal D}\theta}{(2\pi)^{3}} e^{iS\left({\bf X},\theta\right)}
\; ,
\end{eqnarray}
where we sum of all possible transition amplitudes between
points $({\bf X},\theta)$ and $({\bf X}^{\prime},\theta^{\prime})$ of the space $X+\theta$.
Here $N$ is just a normalization constant,
$S\left({\bf X},\theta\right)$ is the action integral
\begin{eqnarray}\label{action}
S\left({\bf X},\theta\right)=\int_{t}^{t^{\prime}}dt^{\prime\prime} \; L(\dot{{\bf X}},\dot{\theta}) \; ,
\end{eqnarray}
and $L$ is the Lagrangian function of the system
\begin{eqnarray}\label{Lagrangianfree}
L({\bf X},\dot{{\bf X}};\theta,\dot{\theta})=\frac{1}{2}m\dot{{\bf X}}^{2}
+\frac{1}{2}\Lambda\dot{\theta}^{2}-V\left({\bf X},\theta\right) \; .
\end{eqnarray}
Here the parameter $\Lambda$ has dimension of mass$^3$ \cite{Amorim1}. As we expected,
the representation of the path integral is given by the functional
integration over the configuration space-$(X+\theta)$ of the
exponential function of the action integral. The function $W$ is a measure
in the $\theta$-integration, and consequently, it attenuates the functional integral
on the $\theta$-space.
Naturally, what emerges in this result is the DFRA-Lagrangian function of the system.
In the next section we apply it to some simple examples, as free particle and the
isotropic harmonic oscillator.

\section{Examples}
\renewcommand{\theequation}{4.\arabic{equation}}
\setcounter{equation}{0}

In this section we will exemplify the formalism, developed before, using two simple systems: the NC free particle and the NC harmonic oscillator.
For the first case we have the propagator as the matrix element (\ref{KXaXb}),
where we insert the completeness relation in the momentum space (\ref{completezarelpkij})
to obtain the integrals
\begin{equation}\label{K0}
K_{0}({\bf x},\theta;{\bf x}^{\prime},\theta^{\prime};\tau)=\langle {\bf X},\theta|e^{-i\hat{H}_{0}\tau}|{\bf X}^{\prime},\theta^{\prime}\rangle= \int \frac{d^{3}{\bf p}}{(2\pi)^{3}}
e^{i{\bf p}\cdot({\bf x}-{\bf x}^{\prime})-\frac{i\tau}{2m}{\bf p}^{2}} \int \frac{d^{3}{\bf k}}{(2\pi)^{3}}
e^{i{\bf k}\cdot(\theta-{\theta}^{\prime})-\frac{i\tau}{2\Lambda}{\bf k}^{2}} \; ,
\end{equation}
where $\hat{H}_{0}$ is the Hamiltonian operator of the free particle in $(x+\theta)$-space (for simplicity we will use from now on $x$ meaning $X$)
\begin{eqnarray}\label{Hlivre}
\hat{H}_{0}(\hat{p}_{i},\hat{\pi}_{ij})=\frac{\hat{p}_{i}^2}{2m}
+\frac{\hat{\pi}_{ij}^{2}}{2\Lambda} \; .
\end{eqnarray}
Using the Gaussian integrals, the free particle propagator is
\begin{equation}\label{GFree}
K_{0}({\bf x}-{\bf x}^{\prime},\theta-\theta^{\prime};t-t^{\prime})=
\frac{i(m\Lambda)^{3/2}}{(t-t^{\prime})^{3}}  \exp\left[\frac{im\left({\bf x}-{\bf x}^{\prime}\right)^{2}+i\Lambda\left(\theta-\theta^{\prime}\right)^{2}}{2(t-t^{\prime})} \right] \; ,
\end{equation}
with the condition $t-t^{\prime}>0$ which is satisfied by causality. We observe that the free propagator is the product between the commutative free propagator in the space-$X$ and the free propagator in the space-$\theta$.

As a second example we have the Hamiltonian of the NC isotropic harmonic oscillator (IHO) \cite{Amorim1}
\begin{eqnarray}\label{HMHS}
\hat{H}_{(IHO)}(\hat{X}_{i},\hat{p}_{i};\hat{\theta}_{ij},\hat{\pi}_{ij})=\frac{\hat{p}_{i}^{2}}{2m}
+\frac{1}{2}m\omega^{2}\hat{X}_{i}^{2}
+\frac{\hat{\pi}_{ij}^{2}}{2\Lambda}
+\frac{1}{2}\Lambda\Omega^{2}\hat{\theta}_{ij}^{2} \; ,
\end{eqnarray}
where $\omega$ and $\Omega$ are the oscillation frequencies in the spaces $({\bf X},\theta)$, respectively.
The operator Hamiltonian (\ref{HMHS}) is the sum of usual operator Hamiltonian of the harmonic oscillator
and NC  Hamiltonian, and consequently, the propagator is the product of usual harmonic oscillator
and the NC  part
\begin{eqnarray}\label{propagadorOHS}
K_{(IHO)}({\bf X},\theta;{\bf X}^{\prime},\theta^{\prime};\tau)=\left(\frac{m
\omega}{2\pi i \sin(\omega\tau)}\right)^{3/2}
\!\!\!\!\exp\left[\frac{im\omega}{2\sin(\omega\tau)} \left(\cos(\omega\tau)
\left({\bf X}^{2}+{\bf X}^{\prime 2} \right)-2{\bf X}\cdot{\bf X}^{\prime} \right)\right]
\nonumber \\
\times \left(\frac{\Lambda\Omega}{2\pi i\sin(\Omega\tau)}\right)^{3/2} \!\!\!\!\exp\left[\frac{i\Lambda\Omega}{2\sin(\Omega\tau)} \left(\cos(\Omega\tau)
\left(\theta^{2}+\theta^{\prime 2} \right)-2{\bf \theta}\cdot{\bf \theta}^{\prime} \right)\right] \; ,
\hspace{1.5cm}
\end{eqnarray}
like in the free propagator case, where $\tau$ is defined as $\tau:=t-t^{\prime}$.
It is easy to write this expression of the propagator in the spectral form (\ref{Kesp}),
and so we obtain the wave function
\begin{eqnarray}\label{wavefIHO}
\Phi_{(n_{1}n_{2}n_{3};\widetilde{n}_{1}\widetilde{n}_{2}\widetilde{n}_{3})}({\bf X},\theta)=\left(\frac{m\omega}{\pi}\right)^{3/4}\prod_{i=1}^{3}e^{-\frac{m\omega}{2}X_{i}^{2}}
\frac{H_{n_{i}}\left( \sqrt{m\omega} X_{i} \right)}{\sqrt{2^{n_{i}}n_{i}!}}
\times \nonumber \\
\times \left(\frac{\Lambda\Omega}{\pi}\right)^{3/4}\prod_{j=1}^{3}e^{-\frac{\Lambda\Omega}{4}\theta_{j}^{2}}
\frac{H_{n_{j}}\left( \sqrt{\Lambda\Omega} \theta_{j} \right)}{\sqrt{2^{n_{j}}n_{j}!}} \; ,
\end{eqnarray}
where $H_{n}$ is the Hermite function, and the energy spectrum
\begin{eqnarray}\label{specIHO}
E_{(n_{1}n_{2}n_{3};\widetilde{n}_{1}\widetilde{n}_{2}\widetilde{n}_{3})}=\sum_{i=1}^{3}\left(n_{i}+\frac{1}{2} \right)\omega
+\sum_{i=1}^{3}\left(\widetilde{n}_{i}+\frac{1}{2} \right)\Omega
\end{eqnarray}
for the NC isotropic harmonic oscillator. For the ground state, we have the wave function
\begin{eqnarray}\label{wavefIHOgstate}
\Phi_{(0;0)}({\bf X},\theta)=\left(\frac{m\omega}{\pi}\right)^{3/4}e^{-\frac{m\omega}{2}{\bf X}^{2}}
\left(\frac{\Lambda\Omega}{\pi}\right)^{3/4}e^{-\frac{\Lambda\Omega}{4}\theta_{ij}\theta^{ij}} \; .
\end{eqnarray}
An important point concerning (\ref{wavefIHOgstate}) is the natural introduction of a weight function $W$
in the $\theta$-sector that appears in the context of NC QFT \cite{Carlson,Morita,Conroy2003}. It must
be connected to $\theta$-integral as an integration measure. We observe explicitly by calculating
the expected value of any function $f$ over the fundamental state
\begin{eqnarray}\label{<f>}
\langle f({\bf X},\theta) \rangle_{0}=\int d^{3}{\bf X} \; d^{3}\theta \; \Phi^{\ast}({\bf X},\theta)
f({\bf X},\theta) \Phi({\bf X},\theta)=
\nonumber \\
=\left(\frac{m\omega}{\pi}\right)^{3/2}\int d^{3}{\bf X} \; e^{-m\omega{\bf X}^{2}} \int d^{3}\theta \; W(\theta)  f({\bf X},\theta) \; ,
\end{eqnarray}
where $W(\theta)$ is
\begin{eqnarray}\label{WthetaMQNC}
W(\theta)=\left(\frac{\Lambda\Omega}{\pi}\right)^{3/2}e^{-\frac{\Lambda\Omega}{2}\theta_{ij}\theta^{ij}} \; .
\end{eqnarray}
We have some properties of the function $W$ [8,12,14,15]
\begin{eqnarray}\label{propertiesWMQNC}
\langle {\bf 1} \rangle_{0} = 1
\hspace{0.5cm} , \hspace{0.5cm}
\langle \theta^{ij} \rangle_{0} = 0
\hspace{0.5cm} , \hspace{0.5cm}
\langle \theta^{ij} \theta^{kl} \rangle_{0} =\frac{\langle \theta^{2} \rangle}{3}\delta^{[ij,kl]} \; \; ,
\end{eqnarray}
with $\langle \theta^{2} \rangle=(2\Lambda\Omega)^{-1}$. More details about the $W$-function will be discussed
in the next section.

The partition function of the NC isotropic harmonic oscillator can be obtained by the usual definition
\begin{eqnarray}\label{PartitionZ}
Z(\beta):={\mbox Tr}\left(e^{-\beta \hat{H}}\right) \; ,
\end{eqnarray}
in which the trace operation is taken over the continuous set of eigenstates
of the position operators $|{\bf X},\theta \rangle$
\begin{eqnarray}\label{Zint}
Z(\beta)=\int d^{3}{\bf X} \; d^{3}\theta \; W(\theta) \; \langle {\bf X},\theta | e^{-\beta \hat{H}} | {\bf X},\theta \rangle \; ,
\end{eqnarray}
where the terms in the propagator we have written ${\bf X}^{\prime}={\bf X}$, $\theta^{\prime}=\theta$
and $\tau=-i\beta$ as the imaginary time interval, so we have
\begin{eqnarray}\label{Zint2}
Z_{(IHO)}(\beta)=\int d^{3}{\bf X} \; d^{3}\theta \; W(\theta) \; K_{(IHO)}\left({\bf X},\theta;{\bf X},\theta,-i\beta \right) \; .
\end{eqnarray}
Substituting the propagator (\ref{propagadorOHS}) and the $W$ function (\ref{WthetaMQNC}) into (4.14),
after a trivial Gaussian integration, we obtain that
\begin{eqnarray}\label{ZOHS}
Z_{(IHO)}(\beta)=\frac{1}{64}\csch^{3}\left(\frac{\omega\beta}{2}\right)\csch^{3}\left(\frac{\Omega\beta}{2}\right)
\left[1+\frac{1}{2}\coth\left(\frac{\Omega\beta}{2}\right)\right]^{-3/2} \; .
\end{eqnarray}
We can see clearly that the NC space contribution is through the $\Omega$ frequency.
The mean energy computed using the partition function is given by the equation
\begin{eqnarray}\label{meanE}
\left. \langle E \rangle =-\frac{\partial }{\partial \beta} \ln Z(\beta) \right|_{\beta=\frac{1}{T}} \; ,
\end{eqnarray}
where the $\beta$ parameter is identified as the inverse of the temperature $T$, and we have the result
\begin{eqnarray}\label{PlanckE}
\langle E \rangle = \frac{3\omega}{2}\coth\left(\frac{\omega}{2T}\right)
+\frac{3\Omega}{2}\left[1+\frac{1}{2}\coth\left(\frac{\Omega}{2T}\right)
-\frac{3/4}{1+\frac{1}{2}\coth\left(\frac{\Omega}{2T}\right)} \right] \; .
\end{eqnarray}
This is the Planck's formula for the average energy of the oscillator. At very low temperatures $(T\ll \omega)$
and $(T\ll \Omega)$, we have the ground state
\begin{eqnarray}\label{EPlanckTpequeno}
\langle E \rangle \approx \frac{3}{2}\left(\omega+\Omega\right) \; ,
\end{eqnarray}
and at high temperature we obtain the classical Boltzman statistics
\begin{eqnarray}\label{EPlanckTgrande}
\langle E \rangle \approx \frac{9T}{2} \; ,
\end{eqnarray}
in which the extra dimension-$\theta$ contributes with a factor of $3/2$.
In the next section we apply the path integral (\ref{intFuncDxDtheta}) to the framework
of quantum field theory in the NC $DFRA$ spacetime discussing the generating
functional, perturbation theory and $n$-points Green functions.

\section{Field theory, Green functions and $\phi^{4}\star$ interaction NC in the DFRA space}
\renewcommand{\theequation}{5.\arabic{equation}}
\setcounter{equation}{0}

Concerning the extension of a NCQFT in DFRA spacetime, the spacetime coordinates $x^{\mu}=(t,{\bf x})$ do not commute with itself satisfying the commutation relation (\ref{xmuxnu}). The parameter $\theta^{\mu\nu}$ is promoted
to be a coordinate of this spacetime.  So, in $D=4$, we have six independents spatial coordinates
associated with $\theta^{\mu\nu}$. The commutation relation of the DFRA algebra in Eqs.
(\ref{xpcom})-(\ref{ppicomm}) can be easily extended
to this NC space
\begin{eqnarray}\label{spacetimealgebraDFRA}
\left[\hat{x}_{\mu},\hat{x}_{\nu}\right] = i\hat{\theta}^{\mu\nu}
\hspace{0.2cm} , \hspace{0.2cm}
\left[\hat{x}_{\mu},\hat{\theta}_{\nu\alpha}\right] = 0
\hspace{0.2cm} , \hspace{0.2cm}
\left[\hat{\theta}_{\mu\nu},\hat{\theta}_{\alpha\beta}\right] = 0
\hspace{0.2cm} , \hspace{0.2cm}
\left[\hat{x}_{\mu},\hat{p}_{\nu} \right] = i\eta_{\mu\nu} \; ,
\nonumber \\
\left[\hat{p}_{\mu},\hat{p}_{\nu} \right] = 0
\hspace{0.2cm} , \hspace{0.2cm}
\left[\hat{p}_{\mu},\hat{\theta}_{\nu\alpha}\right] = 0
\hspace{0.2cm} , \hspace{0.2cm}
\left[\hat{p}_{\mu},\hat{\pi}_{\nu\alpha}\right] = 0
\hspace{0.2cm} , \hspace{0.2cm}
[\hat{x}^{\mu},\hat{\pi}_{\nu\rho}]=-{i\over 2}\delta_{\nu\rho}^{\;\;\;\;\mu\sigma}\hat{p}_{\sigma} \; ,
\end{eqnarray}
where $(\hat{p}_{\mu},\hat{\pi}_{\mu\nu})$ are the momenta operators associated with the coordinates
$(\hat{x}^{\mu},\hat{\theta}^{\mu\nu})$, respectively. The $\theta^{\mu\nu}$ coordinates
are constrained by quantum conditions
\begin{eqnarray}\label{condtheta}
\theta_{\mu\nu}\theta^{\mu\nu}=0
\hspace{0.3cm} \mbox{and} \hspace{0.3cm}
\left(\frac{1}{4}\star\theta_{\mu\nu}\theta^{\mu\nu}\right)^{2}=\lambda_{P}^{8} \; ,
\end{eqnarray}
where $\star\theta_{\mu\nu}=\varepsilon_{\mu\nu\rho\sigma}\theta^{\rho\sigma}$
and $\lambda_{P}$ is the Planck length. In analogy to (\ref{Jij}), the Lorentz group generator
is
\begin{eqnarray}\label{Mmunu}
\hat{M}_{\mu\nu}=\,\hat{X}_{\mu}\hat{p}_{\nu}\,-\,\hat{X}_{\nu}\hat{p}_{\mu}
+\hat{\theta}_{\nu\rho}\hat{\pi}^{\rho}_{\;\mu}-\hat{\theta}_{\mu\rho}\hat{\pi}^{\rho}_{\;\nu} \; ,
\end{eqnarray}
and from (\ref{spacetimealgebraDFRA}) we can write the translations generators
as $\hat{p}_{\mu} = - i \partial_{\mu}\,\,$.
The shifted coordinate operator $\hat{X}^{\mu}$ has the analogous definition of (\ref{X}),
and it satisfies the commutation relations
\begin{eqnarray}\label{Xmupnu}
[\hat{X}_{\mu},\hat{p}_{\nu}]=i\eta_{\mu\nu}
\hspace{0.5cm} \mbox{and} \hspace{0.5cm}
[\hat{X}_{\mu},\hat{X}_{\nu}]=0 \; .
\end{eqnarray}
With these ingredients it is easy to construct the commutation relations
\begin{eqnarray}\label{algebraDFR}
\left[ \hat{p}_\mu , \hat{p}_\nu \right] &=& 0
\hspace{0.2cm} , \nonumber \\
\left[ \hat{M}_{\mu\nu},\hat{p}_{\rho} \right] &=& \,i\,\big(\eta_{\mu\rho}\,\hat{p}_\nu
-\eta_{\mu\nu}\,\hat{p}_\rho\big) \; ,
\hspace{0.1cm} \nonumber \\
\left[\hat{M}_{\mu\nu} ,\hat{M}_{\rho\sigma} \right] &=& i\left(\eta_{\mu\sigma}\hat{M}_{\rho\nu}
-\eta_{\nu\sigma}\hat{M}_{\rho\mu}-\eta_{\mu\rho}\hat{M}_{\sigma\nu}+\eta_{\nu\rho}\hat{M}_{\sigma\mu}\right) \; ,
\end{eqnarray}
and it closes the proper algebra.  We can say that $\hat{p}_\mu$ and $\hat{M}_{\mu\nu}$
are the generators for the extended DFR algebra.
Analyzing the Lorentz symmetry in NCQM following the lines above,
we can introduce a proper theory, for instance, given by a scalar action.
We know, however, that elementary particles are classified according
to the eigenvalues of the Casimir operators of the inhomogeneous Lorentz group.
Hence, let us extend this approach to the Poincar\'e group ${\cal P}$.
Considering the operators presented here, we can in principle consider that
\begin{eqnarray}\label{ElementoG}
\hat{G}={1\over2}\omega_{\mu\nu}\hat{M}^{\mu\nu}
-a^\mu\hat{p}_{\mu}
+{1\over2}b_{\mu\nu}\hat{\pi}^{\mu\nu} \; ,
\end{eqnarray}
is the generator of some group  ${\cal P}'$, which has the Poincar\'e group as a subgroup.
By defining the dynamical transformation of an arbitrary operator $\hat{A}$
in ${\cal H}$ in such a way that $\delta \hat{A}\,=\,i\,[\hat{A},\hat{G}]$
we arrive at the set of transformations,
\begin{eqnarray}\label{transf}
\delta \hat{x}^{\mu}&=&\omega ^\mu_{\,\,\,\,\nu}\hat{x}^{\nu}+a^\mu\nonumber\\
\delta\hat{p}_\mu&=&\omega _\mu^{\,\,\,\,\nu}\hat{p}_\nu\nonumber\\
\delta\hat{\theta}^{\mu\nu}&=&\omega ^\mu_{\,\,\,\,\rho}\hat{\theta}^{\rho\nu}
+ \omega ^\nu_{\,\,\,\,\rho}\hat{\theta}^{\mu\rho}+b^{\mu\nu}\nonumber\\
\delta\hat{\pi}_{\mu\nu}&=&\omega _\mu^{\,\,\,\,\rho}\hat{\pi}_{\rho\nu}
+ \omega _\nu^{\,\,\,\,\rho}\hat{\pi}_{\mu\rho}\nonumber\\
\delta \hat{M}_1^{\mu\nu}&=&\omega ^\mu_{\,\,\,\,\rho}\hat{M}_1^{\rho\nu}
+ \omega ^\nu_{\,\,\,\,\rho}\hat{M}_1^{\mu\rho}+a^\mu\hat{p}^\nu-a^\nu\hat{p}^\mu\nonumber\\
\delta \hat{M}_2^{\mu\nu}&=&\omega ^\mu_{\,\,\,\,\rho}\hat{M}_2^{\rho\nu}
+ \omega ^\nu_{\,\,\,\,\rho}\hat{M}_{2}^{\mu\rho}+b^{\mu\rho}\hat{\pi}_\rho^{\,\,\,\,\nu}
+ b^{\nu\rho}\hat{\pi}_{\,\,\,\rho}^{\mu}\nonumber\\
\delta \hat{x}^{\mu}&=&\omega ^\mu_{\,\,\,\,\nu}\hat{x}^{\nu}+a^\mu+{1\over2}b^{\mu\nu}\hat{p}_{\nu} \; .
\end{eqnarray}

We observe that there is an unexpected term in the last equation of (\ref{transf}).
This is a consequence of the coordinate operator in (\ref{X}), which is a nonlinear combination
of operators that act on different Hilbert spaces.

The action of ${\cal P}'$ upon Hilbert space operators is in some sense equal to the action of the
Poincar\'e group with an additional translation operation on the ($\hat{\theta}^{\mu\nu}$) sector.
Its generators, all of them, close in a commutation algebra. Hence, ${\cal P}'$ is a well defined
group of transformations.
As a matter of fact, the commutation of two transformations closes in the algebra
\begin{equation}\label{algebray}
[\delta_2,\delta_1]\,\hat{y}=\delta_3\,\hat{y} \; ,
\end{equation}
where ${\mathbf y}$ represents any one of the operators appearing in (\ref{transf}).
The parameters composition rule is given by
\begin{eqnarray}\label{DecompOmega}
\omega^\mu_{3\,\,\nu}&=&\omega^\mu_{1\,\,\,\,\alpha}\omega^\alpha_{2\,\,\,\,\nu}-\omega^\mu_{2\,\,\,\,\alpha}\omega^\alpha_{1\,\,\,\,\nu}\nonumber\\
a_3^\mu&=&\omega^\mu_{1\,\,\,\nu}a_2^\nu-\omega^\mu_{2\,\,\,\nu}a_1^\nu \nonumber  \\
b_3^{\mu\nu}&=&\omega^\mu_{1\,\,\,\rho}b_2^{\rho\nu}-\omega^\mu_{2\,\,\,\rho}b_1^{\rho\nu}-\omega^\nu_{1\,\,\,\rho}b_2^{\rho\mu}+
\omega^\nu_{2\,\,\,\rho}b_{1}^{\rho\mu} \,\,.
\end{eqnarray}
To sum up, the framework showed above demonstrated that in NCQM, the physical coordinates
do not commute and the respective eigenvectors cannot be used to form a basis
in ${\cal H}={\cal H}_1 \oplus {\cal H}_2$ \cite{Amorim4}.  This can be accomplished
with the Bopp shift defined in
(\ref{X}) with (\ref{Xmupnu}) as consequence.
So, we can introduce a coordinate basis
$|X^{\prime}, \theta^{\prime} \rangle = |X^{\prime}\rangle \otimes |\theta^{\prime}\rangle$
and $|p^{\prime}, k^{\prime} \rangle = |p^{\prime}\rangle \otimes |k^{\prime}\rangle$,
in such a way that
\begin{eqnarray}\label{eqeigenvaluesXtheta}
\hat{X}^{\mu}|X^{\prime}, \theta^{\prime}\rangle=X^{\prime\mu}|X^{\prime}, \theta^{\prime}\rangle
\qquad \mbox{and} \qquad \hat{\theta}^{\mu\nu}|X^{\prime},
\theta^{\prime}\rangle=\theta^{\prime\mu\nu}|X^{\prime},\theta^{\prime}\rangle \; ,
\end{eqnarray}
and
\begin{eqnarray}\label{eqautovalorespmupimunu}
\hat{p}_{\mu}|p^{\prime},k^{\prime} \rangle
= p^{\prime}_{\mu}|p^{\prime},k^{\prime} \rangle
\hspace{0.5cm} , \hspace{0.5cm} \quad
\hat{\pi}_{\mu\nu}|p^{\prime},k^{\prime} \rangle
= k^{\prime}_{\mu\nu}|p^{\prime},k^{\prime} \rangle \; .
\end{eqnarray}

The wave function $\phi(X^{\prime}, \theta^{\prime})=\langle X^{\prime}, \theta^{\prime}|\phi\rangle$
represents the physical state $|\phi\rangle$ in the coordinate basis defined above.
This wave function satisfies some wave equation that can be derived from an action,
through a variational principle, as usual. In \cite{Amorim4}, the author constructed directly
an ordinary relativistic free quantum theory.
It was assumed that the physical states are annihilated
by the mass-shell condition
\begin{eqnarray}\label{reldisp}
\left(\hat{p}_{\mu}\hat{p}^{\mu}-m^2\right)|\phi \rangle = 0 \; ,
\end{eqnarray}
demonstrated through the Casimir operator $C_{1}=\hat{p}_{\mu}\hat{p}^{\mu}$ (for more algebraic details
see \cite{Amorim4}).
It is easy to see that in the coordinate representation, this originates the NC KG equation.
Condition (\ref{reldisp}) selects the physical states that must be invariant under gauge transformations.
To treat the NC case, let us assume that the second mass-shell condition
\begin{eqnarray}\label{reldisppi}
\left( \hat{\pi}_{\mu\nu}\hat{\pi}^{\mu\nu}-\Delta \right)|\phi\rangle = 0 \; ,
\end{eqnarray}
and must be imposed on the physical states, where $\Delta$ is some constant with dimension $M^4$,
which sign and value can be defined if $\pi$ is spacelike, timelike or null.
Analogously the Casimir invariant is $C_{2}=\hat{\pi}_{\mu\nu}\hat{\pi}^{\mu\nu}$,
demonstrated the validity of (\ref{reldisp}) (see \cite{Amorim4} for details).

Both equations (\ref{reldisp}) and ({\ref{reldisppi}) permit us to construct a general
expression for the plane wave solution such as \cite{Amorim4}
\begin{eqnarray}\label{phiXtheta}
\phi(x^{\prime},\theta^{\prime}):= \langle X^{\prime},\theta^{\prime}|\phi\rangle
=\int \frac{d^{4}p}{(2\pi)^{4}} \frac{d^{6}k}{\left(2\pi\lambda^{-2}\right)^{6}} 
\, \widetilde{\phi}(p,k^{\mu\nu})
\, \exp \left( ip_{\mu}x^{\prime\mu}+\frac i2 k_{\mu\nu} \theta^{\prime \mu\nu} \right) \; ,
\end{eqnarray}
where $p^2\,-\,m^2=0$ and $k^2\,-\,\Delta=0$, and we have used that $p\cdot X=p \cdot x$.
The length $\lambda^{-2}$ is introduced conveniently in the $k$-integration 
in order to keep the usual dimensions of the fields since the action $S$ must have null dimension in natural units.
Or an even weight function that will be defined in a few moments,
used to make the bridge between the formalism in $D=4+6$ and the standard one in $D=4$.
Or finally it can be understood as a distribution used to impose further conditions \cite{DFR}.

In coordinate representation, the operators $(\hat{p},\hat{\pi})$ are written in terms of the
derivatives
\begin{eqnarray}\label{repcoordinateppi}
\hat{p}_{\mu} \rightarrow -i\p_{\mu}
\hspace{0.6cm} \mbox{and} \hspace{0.6cm}
\hat{\pi}_{\mu\nu} \rightarrow -i\frac{\p}{\p \theta^{\mu\nu}} \; ,
\end{eqnarray}
and consequently, both (\ref{reldisp}) and (\ref{reldisppi}) are just the
Klein-Gordon equations
\begin{eqnarray}\label{EqKleinGordonp}
\left( \Box \,+\,M^2 \right) \phi (x, \theta )\,=\,0
\end{eqnarray}
and
\begin{eqnarray}\label{EqKleinGordonpi}
(\Box_{\theta}\,+\,\Delta^4 ) \, \phi (x, \theta)\,=\,0 \; ,
\end{eqnarray}
respectively, where we have defined $\Box_{\theta}=\frac 12\,\p^{\mu\nu}\,\p_{\mu\nu}$
and $\p_{\mu\nu}=\frac{\p}{\p \theta^{\prime\mu\nu}}$, with $\eta^{\mu\nu}=\mbox{diag}(1,-1,-1,-1)$.
It can be summarized in a single equation for the massive scalar field $\phi$
\begin{eqnarray}\label{NCKG}
\left(\Box +\lambda^2\Box_\theta+m^2\right)\phi(x,\theta)=0 \; ,
\end{eqnarray}
which is the Klein-Gordon equation in DFRA space. The parameters $M$ and $\Delta$ have mass dimension,
and it is related with the mass scalar field, that is, $m^{2}=M^{2}+\lambda^{2}\Delta^{4}$.
Substituting the wave plane solution (\ref{phiXtheta}), we obtain the mass invariant
\begin{eqnarray}\label{MassInv}
p^{2}+\frac{\lambda^{2}}{2}k_{\mu\nu}k^{\mu\nu}=m^2 \; .
\end{eqnarray}
where $\lambda$ is a parameter with dimension of length defined before,
as the Planck length. We define the components of the $k$-momentum
$k^{\mu\nu}=(-{\bf k},-\widetilde{{\bf k}})$ and $k_{\mu\nu}=({\bf k},\widetilde{{\bf k}})$,
to get the DFRA dispersion relation
\begin{eqnarray}\label{RelDispDFRA}
\omega({\bf p},{\bf k},\widetilde{{\bf k}})=\sqrt{{\bf p}^{2}
+\lambda^{2}\left({\bf k}^{2}+\widetilde{{\bf k}}^{2}\right)+m^2} \; ,
\end{eqnarray}
in which $\widetilde{k}_{i}$ is the dual vector of the components $k_{ij}$,
that is, $k_{ij}=\epsilon_{ijk}\widetilde{k}_{k}$ $(i,j,k=1,2,3,)$.

To propose the action of a scalar field we need to define the Weyl
representation for DFRA operators. It is given by the mapping
\begin{eqnarray}\label{mapweyl}
\hat{{\cal W}}(f)(\hat{x},\hat{\theta})=\int\frac{d^{4}p}{(2\pi)^{4}}
\frac{d^{6}k}{\left(2\pi\lambda^{-2}\right)^{6}} \;
\widetilde{f}(p,k^{\mu\nu}) \; e^{ip \cdot \hat{x}+\frac{i}{2} k \cdot \hat{\theta}} \; ,
\end{eqnarray}
where $(\hat{x},\hat{\theta})$ are position operators satisfying the DFRA algebra, $p^{\mu}$ and $k^{\mu\nu}$
are the conjugated momentum of the coordinates $x^{\mu}$ and $\theta^{\mu\nu}$ , respectively.
The Weyl symbol provides a map from the operator algebra to the algebra of functions
equipped with a star-product, via the Weyl-Moyal correspondence
\begin{eqnarray}\label{WeylMoyal}
\hat{f}(\hat{x},\hat{\theta}) \; \hat{g}(\hat{x},\hat{\theta})
\hspace{0.3cm} \leftrightarrow \hspace{0.3cm}
f(x,\theta) \star g(x,\theta) \; ,
\end{eqnarray}
and the star-product turns out to be the same as in the usual NC case
\begin{eqnarray}\label{ProductMoyal}
\left. f(x,\theta) \star g(x,\theta) =
e^{\frac{i}{2}\theta^{\mu\nu}\partial_{\mu}\partial^{\prime}_{\nu}}
f(x,\theta) g(x^{\prime},\theta) \right|_{x^{\prime}=x} \; ,
\end{eqnarray}
for any functions $f$ and $g$. The Weyl operator (\ref{mapweyl}) has the following trace properties
\begin{eqnarray}\label{traceW}
\mbox{Tr}\left[\hat{{\cal W}}(f)\right]=\int d^{4}x \; d^{6}\theta \; W(\theta) \; f(x,\theta) \; ,
\end{eqnarray}
and for a product of $n$ functions $(f_{1},...,f_{n})$
\begin{eqnarray}\label{traceWfs}
\mbox{Tr}\left[ \hat{{\cal W}}(f_{1}) ... \hat{{\cal W}}(f_{n}) \right]=
\int d^{4}x \; d^{6}\theta \; W(\theta) \; f_{1}(x,\theta) \star ... \star f_{n}(x,\theta) \; .
\end{eqnarray}

The function $W$ is a Lorentz invariant integration-$\theta$ measure.
This weight function is introduced in the context of NC field theory to control divergences of the integration
in the $\theta$-space \cite{Carlson,Morita,Conroy2003}. It will permits us to work with series expansions
in $\theta$, {\it i.e.}, with truncated power series expansion of functions of $\theta$.
For any large $\theta_{\mu\nu}$ it falls to zero quickly so that all integrals are well defined,
in that it is assumed the normalization condition
\begin{eqnarray}\label{normalizationW}
\langle {\bf 1} \rangle = \int d^{6}\theta \; W(\theta) = 1 \; .
\end{eqnarray}
The function $W$ should be a even function of
$\theta$, that is, $W(-\theta)=W(\theta)$,
and consequently it implies that
\begin{eqnarray}\label{intWtheta}
\langle \theta^{\mu\nu} \rangle = \int d^{6} \theta \; W(\theta) \; \theta^{\mu\nu} = 0 \; .
\end{eqnarray}
The non trivial integrals are expressed in terms of the invariant
\begin{eqnarray}\label{theta2n}
\langle \theta^{2 n} \rangle=\int d^{6}\theta \; W(\theta) \; (\theta_{\mu\nu}\theta^{\mu\nu})^{n}
\; , \; \mbox{with} \; n \in {\mathbb Z}_{+} \; ,
\end{eqnarray}
in which the normalization condition corresponding to the case $n=0$. For $n=1$, we have
\begin{eqnarray}\label{intWtheta2}
\int d^{6}\theta \; W(\theta) \; \theta^{\mu\nu} \theta^{\rho\lambda}
=\frac{\langle \theta^{2} \rangle}{6}{\bf 1}^{[\mu\nu,\rho\lambda]} \; ,
\end{eqnarray}
where ${\bf 1}^{[\mu\nu,\rho\lambda]}:=(g^{\mu\rho}g^{\nu\lambda}-g^{\mu\lambda}g^{\nu\rho})/2$
is the identity antisymmetric on index $(\mu\nu)$ and $(\rho\lambda)$. For $n=2$, we have
\begin{equation}\label{intWtheta4}
\int d^{6}\theta \; W(\theta) \; \theta^{\mu\nu} \theta^{\rho\lambda} \theta^{\alpha\beta} \theta^{\gamma\sigma}
=\frac{\langle \theta^{4} \rangle}{48}\left({\bf 1}^{[\mu\nu,\rho\lambda]}{\bf 1}^{[\alpha\beta,\gamma\sigma]}
+{\bf 1}^{[\mu\nu,\alpha\beta]}{\bf 1}^{[\rho\lambda,\gamma\sigma]} + {\bf 1}^{[\mu\nu,\gamma\sigma]}
{\bf 1}^{[\rho\lambda,\alpha\beta]}  \right) \; .
\end{equation}
A explicit form for the weight function $W$ that satisfies all this previous properties is
\begin{eqnarray}\label{formWtheta}
W(\theta)=\left(\frac{1}{2\pi \lambda^{4}}\right)^{3}
e^{-\frac{\left|\theta_{\mu\nu}\theta^{\mu\nu}\right|}{4\lambda^{4}}} \; ,
\end{eqnarray}
and the absolute value has been introduced to assure that there is not directions
in which $W$-function blows up to infinity. In the limit $\lambda \rightarrow 0$
the $W$-function tends to a Dirac's delta function, that when integrated in $d^{6}\theta$,
it has the interpretation of spatial volume of the extra dimension-$\theta$.
Using the previous properties we have an important integration for us
\begin{eqnarray}\label{IntWthetaexp}
\int d^{6}\theta \; W(\theta) \; e^{\frac{i}{2} k_{\mu\nu} \theta^{\mu\nu}} =
e^{-\frac{\lambda^{4}}{4}\left|k_{\mu\nu}k^{\mu\nu}\right|} \; ,
\end{eqnarray}
%
With the definition of the Moyal product (\ref{ProductMoyal}) it is
trivial to obtain the property
\begin{eqnarray}\label{Idmoyalproduct2}
\int d^{4}x\,d^{6}\theta \; W(\theta) \; f(x,\theta) \star g(x,\theta)=
\int d^{4}x\,d^{6}\theta \; W(\theta) \; f(x,\theta) \; g(x,\theta) \; .
\end{eqnarray}
%
The physical interpretation of the average of the components of $\theta^{\mu\nu}$, {\it i.e.}
$\langle \theta^{2} \rangle$, is the definition of the NC  energy scale \cite{Carlson}
\begin{eqnarray}\label{LambdaNC}
\Lambda_{NC}=\left(\frac{12}{\langle \theta^{2} \rangle} \right)^{1/4}=:\frac{1}{\lambda} \; ,
\end{eqnarray}
in which $\lambda$ is the fundamental length scale
which appeared in the KG equation (\ref{NCKG}), and in the dispersion relation
(\ref{RelDispDFRA}). This approach has the advantage of being independent of the
form of the function $W$, at least for lowest-order processes.
The study of Lorentz-invariant NC  QED,
as Bhabha scattering, dilepton and diphoton production
to LEP data led the authors of \cite{Conroy2003,Carone} to the bound
\begin{eqnarray}\label{boundLambda}
\Lambda_{NC} > 160 \; GeV \; \; 95 \% \; C.L. \; .
\end{eqnarray}

After the discussion of the $W$-function it can be postulated the completeness relations
\begin{eqnarray}\label{completezarelXtheta}
\int d^{4}x^{\prime} d^{6}\theta^{\prime} \, W(\theta^{\prime}) |X^{\prime},\theta^{\prime} \rangle
\langle X^{\prime},\theta^{\prime} | ={\bf 1} \; ,
\end{eqnarray}
and
\begin{eqnarray}\label{completezarelpkmunu}
\int \frac{d^{4}p^{\prime}}{(2\pi)^{4}} \; \frac{d^{6}k^{\prime}}{(2\pi\lambda^{-2})^{6}} 
\; |p^{\prime},k^{\prime} \rangle
\langle p^{\prime},k^{\prime} | ={\bf 1} \; .
\end{eqnarray}
Using the previous completeness relations and the integral (\ref{IntWthetaexp}), we obtain
\begin{eqnarray}\label{ortognalityrelationXtheta}
\langle X^{\prime},\theta^{\prime} |X^{\prime\prime},\theta^{\prime\prime} \rangle =
\delta^{(4)}\!\left(x^{\prime}-x^{\prime\prime}\right)W^{-1}(\theta^{\prime})\delta^{(6)}\!
\left(\theta^{\prime}-{\bf \theta}^{\prime\prime}\right) \; ,
\end{eqnarray}
and
\begin{eqnarray}\label{ortognalityrelationppi}
\langle p, k |p^{\prime}, k^{\prime} \rangle =
(2\pi)^{4}\delta^{(4)}\!\left(p-p^{\prime}\right)
e^{-\frac{\lambda^{4}}{4}\left|\left(k_{\mu\nu}-k_{\mu\nu}^{\prime}\right)^{2}\right|} \; ,
\end{eqnarray}
where we have used $\langle \theta^{2} \rangle=12\lambda^{4}$ of (\ref{LambdaNC}),
and the matrix elements
\begin{eqnarray}\label{XthetapmuXtheta}
\langle X^{\prime},\theta^{\prime}|\hat{p}_{\mu}| X^{\prime\prime},\theta^{\prime\prime} \rangle \,=
\,-\,i\,\partial^{\prime}_{\mu}\,\delta^{(4)}(x^{\prime}-x^{\prime\prime})
W^{-1}(\theta^{\prime})\delta^{(6)}(\theta^{\prime}-\theta^{\prime\prime})
\; ,
\end{eqnarray}
and
\begin{eqnarray}\label{XthetapimunuXtheta}
\langle X^{\prime},\theta^{\prime}|\hat{\pi}_{\mu\nu}| X^{\prime\prime},\theta^{\prime\prime}\rangle \,
=\delta^{(4)}(x^{\prime}-x^{\prime\prime})(\,-\,i\,)
\frac{\p}{\p \theta^{\prime \mu\nu}}\left( W^{-1}(\theta^{\prime})
\delta^{(6)}(\theta^{\prime}-\theta^{\prime\prime})  \right) \; ,
\end{eqnarray}
that confirms the differential representation of (\ref{repcoordinateppi}).
The result (\ref{ortognalityrelationppi}) reveals that the
canonical momentum $k^{\mu\nu}$ associated to $\theta^{\mu\nu}$ is not conserved
due to introduction of the $W$-function.

Since we constructed the NC KG equation, we will now provide its correspondent action.
We use the definition of Moyal-product (\ref{ProductMoyal}) to write the action with a $\phi^{4}\star$
interaction term
\begin{equation}\label{actionscalarstar}
S(\phi)=\int d^{4}x \,d^{6}\theta \, W(\theta) \left( \frac{1}{2} \partial_{\mu}\phi \star \partial^{\mu}\phi   +
{{\lambda^2}\over4} \partial_{\mu\nu}\phi \star \partial^{\mu\nu}\phi
-\frac{1}{2}m^2\phi \star \phi -\frac{g}{4!} \phi \star \phi \star \phi \star \phi \right) \; ,
\end{equation}
where $g$ is a constant coupling, and using the identity (\ref{Idmoyalproduct2}),
the free part of the action is reduced to the usual product one
\begin{eqnarray}\label{actionscalar}
S(\phi)=\int d^{4}x \,d^{6}\theta \, W(\theta) \left[ \frac{1}{2} \left(\partial_{\mu}\phi\right)^{2} +
{{\lambda^2}\over 4} \left(\partial_{\mu\nu}\phi\right)^{2}
-\frac{1}{2}m^2\phi^{2}-\frac{g}{4!} \left(\phi \star \phi\right)^{2}  \right] \; ,
\end{eqnarray}
where $W$ is the previous weight function in the measure-$\theta$,
that can be used to connect $D=4+6$ and $D=4$ DFRA formalisms.
Firstly, we study the free part of (\ref{actionscalar}) by revisiting
the retarded, advanced and causal free Green functions \cite{EMCAbreuMJNeves2011}.

The NC KG equation DFRA in the presence of an external source $J$ admits two solutions
of the type
\begin{eqnarray}\label{solutionphiin}
\phi(x,\theta)=\phi_{in}(x,\theta)
+\!\!\int d^{4}x^{\prime} d^{6}\theta^{\prime} \,W(\theta^{\prime})\;
\Delta^{(-)}(x-x^{\prime},\theta-\theta^{\prime}) J(x^{\prime},\theta^{\prime}) \; ,
\end{eqnarray}
and
\begin{eqnarray}\label{solutionphiin}
\phi(x,\theta)=\phi_{out}(x,\theta)
+\!\!\int d^{4}x^{\prime} d^{6}\theta^{\prime} \,W(\theta^{\prime})\;
\Delta^{(+)}(x-x^{\prime},\theta-\theta^{\prime}) J(x^{\prime},\theta^{\prime}) \; ,
\end{eqnarray}
in which $\phi_{in}$ and $\phi_{out}$ are asymptotic fields and solutions of the NC KG free equation.
The $\Delta^{(-)}$ and $\Delta^{(+)}$ are the retarded and advanced Green functions
in the NC DFRA, respectively. It is obtained inverting the Green equation
\begin{eqnarray}\label{EqGreen}
\left(\Box+\lambda^{2}\Box_{\theta}+m^{2}\right)\Delta^{(\pm)}(x-x^{\prime};\theta-\theta^{\prime})
=\delta^{(4)}(x-x^{\prime})W^{-1}(\theta)\delta^{(6)}(\theta-\theta^{\prime}) \; ,
\end{eqnarray}
by the traditional Fourier method, we have
\begin{equation}\label{GRetAdv}
\Delta^{(\mp)}(x-x^{\prime};\theta-\theta^{\prime})=\int_{(\mp)}
\frac{dp_{0}}{2\pi} \frac{d^{3}{\bf p}}{(2\pi)^{3}} \int \frac{d^{6}k}{(2\pi\lambda^{-2})^{6}}
\,\frac{e^{ip\cdot\left(x-x^{\prime}\right)+
\frac{i}{2}k_{\mu\nu}\left(\theta-\theta^{\prime}\right)^{\mu\nu}}}
{(p_{0} \pm i\varepsilon)^2-{\bf p}^{2}+\frac{\lambda^{2}}{2}k_{\mu\nu}k^{\mu\nu}-m^{2}} \; .
\end{equation}
Here we have added the prescription $p_{0} \mapsto \omega + i\varepsilon$ (retarded Green function),
and $p_{0} \mapsto \omega - i\varepsilon$ (advanced Green function). The symbols $(\mp)$
in the $p^{0}$-integration denotes the convenient contour associated to those Green functions.
For more details on these NC Green functions, see \cite{EMCAbreuMJNeves2011}.

For the formalism of perturbative field theory we need the Causal Green function,
which has a different prescription comparing with (\ref{GRetAdv}). To simplify
our task we write the free part of (\ref{actionscalar}) in the momentum
space by using the Fourier representation (\ref{phiXtheta}) and the integral
(\ref{IntWthetaexp}), so we obtain
\begin{eqnarray}\label{actionscalark}
S_{0}(\widetilde{\phi})=\int \frac{d^{4}p}{(2\pi)^{4}} \, \frac{d^{6}k}{(2\pi\lambda^{-2})^{6}} \,
\frac{d^{6}k^{\prime}}{(2\pi\lambda^{-2})^{6}} \, 
e^{-\frac{\lambda^{4}}{4}\left|\left(k_{\mu\nu}+k_{\mu\nu}^{\prime}\right)^{2}\right|} \, \times
\nonumber \\
\times \widetilde{\phi}(-p,k_{\mu\nu}^{\prime}) \frac{1}{2} \left( p^{2} +
{{\lambda^2}\over2} k_{\mu\nu}^{\prime}k^{\mu\nu}-m^2  \right) \widetilde{\phi}(p,k_{\mu\nu}) \; .
\end{eqnarray}
Therefore we postulate the free generating functional defined with configurations in the momentum space
\begin{eqnarray}\label{Zfree}
Z_{0}(\widetilde{J})=\int {\cal D}\widetilde{\phi} \, \exp
\left[ iS_{0}(\widetilde{\phi};-i\varepsilon)+
\hspace{3cm} \right. \nonumber \\
\left. +i\int \frac{d^{4}p}{(2\pi)^{4}} \, \frac{d^{6}k}{(2\pi\lambda^{-2})^{6}} \,
\frac{d^{6}k^{\prime}}{(2\pi\lambda^{-2})^{6}} \,
\,e^{-\frac{\lambda^{4}}{4}\left|\left(k_{\mu\nu}+k_{\mu\nu}^{\prime}\right)^{2}\right|} 
\widetilde{J}(-p,k_{\mu\nu}^{\prime})
\widetilde{\phi}(p,k_{\mu\nu}) \right] \; ,
\end{eqnarray}
where the prescription $m^{2} \mapsto m^{2}-i\varepsilon$ has been added to the mass term
in the free action (\ref{actionscalark}) to obtain a functional integration well defined. This
prescription gives rise the causal Green function. To obtain the Feynman propagator
(causal Green function) we make the transformation
$\widetilde{\phi} \mapsto \widetilde{\phi} - (p^{2} +
{{\lambda^2}\over4} k_{\mu\nu}^{\prime}k^{\mu\nu}-m^2+i\varepsilon)^{-1}\widetilde{J}$,
and the $Z_{0}$ functional is
\begin{equation}\label{Z0J}
Z_{0}(\widetilde{J})=\exp\left[-\frac{1}{2} \int \frac{d^{4}p}{(2\pi)^{4}} \, \frac{d^{6}{k}}{(2\pi\lambda^{-2})^{6}} \,
\frac{d^{6}{k}^{\prime}}{(2\pi\lambda^{-2})^{6}} \,\widetilde{J}(-p,k_{\mu\nu}^{\prime})
\frac{i\,e^{-\frac{\lambda^{4}}{4}\left|\left(k_{\mu\nu}+k_{\mu\nu}^{\prime}\right)^{2}\right|}}
{p^{2}+\frac{\lambda^{2}}{2}k_{\mu\nu}^{\prime}k^{\mu\nu}-m^{2}+i\varepsilon}
\widetilde{J}(p,k_{\mu\nu}) \right] \; .
\end{equation}
The free $2$-points Green function in the momentum space is defined by derivatives
of $Z_{0}$ in relation to $\widetilde{J}$, so we have
\begin{eqnarray}\label{Deltapk}
\widetilde{\Delta}_{F}^{(2)}(p, p^{\prime};k_{\mu\nu}, k_{\mu\nu}^{\prime})
= (2\pi)^{4}\delta^{(4)}(p+p^{\prime})
\frac{ie^{-\frac{\lambda^{4}}{4}\left|\left(k_{\mu\nu}+k_{\mu\nu}^{\prime}\right)^{2}\right|}}
{p^{2}+\frac{\lambda^{2}}{2}k_{\mu\nu}^{\prime}k^{\mu\nu}-m^{2}+i\varepsilon} \; ,
\end{eqnarray}
where the causal Green function in the $(x+\theta)$-space is the Fourier transform
\begin{eqnarray}\label{GFourier}
\Delta_{F}(x-x^{\prime};\theta,\theta^{\prime})&=&\int_{(F)} \! \frac{dp_{0}}{2\pi}
\frac{d^{3}{\bf p}}{(2\pi)^{3}} \; e^{ip\cdot(x-x^{\prime})} \times \nonumber \\
&\times&\int \frac{d^{6}{k}}{(2\pi\lambda^{-2})^{6}}
\frac{d^{6}{k}^{\prime}}{(2\pi\lambda^{-2})^{6}} \,
e^{\frac{i}{2}\left(k_{\mu\nu}\theta^{\mu\nu}-k_{\mu\nu}^{\prime}\theta^{\prime\mu\nu}\right)}
\frac{ie^{-\frac{\lambda^{4}}{4}\left|\left(k_{\mu\nu}+k_{\mu\nu}^{\prime}\right)^{2}\right|}}{p_{0}^{2}-{\bf p}^{2}
+\frac{\lambda^{2}}{2}k_{\mu\nu}^{\prime}k^{\mu\nu}-m^{2}+i\varepsilon} \; , \nonumber \\
\end{eqnarray}
where the symbol $(F)$ indicates the integration contour correspondent to Causal Green function.

For the interaction part $\phi^{4}\star$ of (\ref{actionscalar}) we also write it
in the momentum space by using the Fourier transform (\ref{phiXtheta}), the interaction is
\begin{eqnarray}\label{actionintmomentumdthetaw}
S_{int}(\widetilde{\phi})=-\frac{g}{4!}\int \prod_{i=1}^{4} \frac{d^{4}p_{i}}{(2\pi)^{4}}
\frac{d^{6}{k}_{i}}{(2\pi\lambda^{-2})^{6}} \; \widetilde{\phi}(p_{i},{k}_{i\mu\nu})
(2\pi)^{4} \delta^{(4)}\left(p_{1}+p_{2}+p_{3}+p_{4} \right) \times
\nonumber \\
\times \int d^{6}\theta \, W(\theta)
F(p_{1},p_{2},p_{3},p_{4}) \,
e^{\frac{i}{2}\left(k_{1\mu\nu}+k_{2\mu\nu}+k_{3\mu\nu}+k_{4\mu\nu}\right)\theta^{\mu\nu}}  \; ,
\end{eqnarray}
where $F$ is a function totally symmetric exchanging the momenta $(p_{1},p_{2},p_{3},p_{4})$
\begin{eqnarray}\label{F}
F(p_{1},p_{2},p_{3},p_{4})=\frac{1}{3}\left[ \cos \left(\frac{p_{1} \wedge p_{2} }{2}\right) \cos \left(\frac{p_{3} \wedge p_{4} }{2}\right)
+\cos \left(\frac{p_{1} \wedge p_{3} }{2}\right)\cos \left(\frac{p_{2} \wedge p_{4} }{2}\right)
\right. \nonumber \\
\left. +\cos \left(\frac{p_{1} \wedge p_{4} }{2}\right) \cos \left(\frac{p_{2} \wedge p_{3} }{2}\right) \right] \; ,
\end{eqnarray}
and the symbol $\wedge$ means the product $p_{i} \wedge p_{j}=\theta^{\mu\nu}p_{\mu i} p_{\nu j}$
if $i \neq j$ $(i,j=1,2,3,4)$, and $p_{i} \wedge p_{j}=0$, if $i=j$. The $\theta$-integral of
(\ref{actionintmomentumdthetaw}) can be calculated with the help of (\ref{IntWthetaexp})
which, after some manipulations we obtain
\begin{eqnarray}\label{actionintmomentum}
S_{int}(\widetilde{\phi})=-\frac{g}{12 \times 4!}\int \prod_{i=1}^{4} \frac{d^{4}p_{i}}{(2\pi)^{4}}
\frac{d^{6}{k}_{i}}{(2\pi\lambda^{-2})^{6}} \; \widetilde{\phi}(p_{i},k_{i\mu\nu})
(2\pi)^{4} \delta^{(4)}\left(p_{1}+p_{2}+p_{3}+p_{4} \right) \times
\hspace{0.6cm}\nonumber \\
\left[e^{-\frac{\lambda^{4}}{4}\left(k_{s\mu\nu}+p_{1\mu}p_{2\nu}+p_{3\mu}p_{4\nu}\right)^{2}}
+e^{-\frac{\lambda^{4}}{4}(k_{s\mu\nu}-p_{1\mu}p_{2\nu}-p_{3\mu}p_{4\nu})^{2}}
+e^{-\frac{\lambda^{4}}{4}(k_{s\mu\nu}+p_{1\mu}p_{2\nu}-p_{3\mu}p_{4\nu})^{2}}
\right. \nonumber \\
\left.
+e^{-\frac{\lambda^{4}}{4}(k_{s\mu\nu}-p_{1\mu}p_{2\nu}+p_{3\mu}p_{4\nu})^{2}}
+e^{-\frac{\lambda^{4}}{4}(k_{s\mu\nu}+p_{1\mu}p_{3\nu}+p_{2\mu}p_{4\nu})^{2}}
+e^{-\frac{\lambda^{4}}{4}(k_{s\mu\nu}-p_{1\mu}p_{3\nu}-p_{2\mu}p_{4\nu})^{2}}
\right. \nonumber \\
\left.
+e^{-\frac{\lambda^{4}}{4}(k_{s\mu\nu}+p_{1\mu}p_{3\nu}-p_{2\mu}p_{4\nu})^{2}}
+e^{-\frac{\lambda^{4}}{4}(k_{s\mu\nu}-p_{1\mu}p_{3\nu}+p_{2\mu}p_{4\nu})^{2}}
+e^{-\frac{\lambda^{4}}{4}(k_{s\mu\nu}+p_{1\mu}p_{4\nu}+p_{2\mu}p_{3\nu})^{2}}
\right. \nonumber \\
\left.
+e^{-\frac{\lambda^{4}}{4}(k_{s\mu\nu}-p_{1\mu}p_{4\nu}-p_{2\mu}p_{3\nu})^{2}}
+e^{-\frac{\lambda^{4}}{4}(k_{s\mu\nu}+p_{1\mu}p_{4\nu}-p_{2\mu}p_{3\nu})^{2}}
+e^{-\frac{\lambda^{4}}{4}(k_{s\mu\nu}-p_{1\mu}p_{4\nu}+p_{2\mu}p_{3\nu})^{2}}
\right] .
\end{eqnarray}
The vertex $V^{(4)}$ of the $\phi^{4}\star$ interaction in the momentum space is
\begin{eqnarray}\label{VertexV4}
V^{(4)}(p_{1},...,p_{4};k_{1\mu\nu},...,k_{4\mu\nu})=-\frac{g}{6}
(2\pi)^{4} \delta^{(4)}\left(p_{1}+p_{2}+p_{3}+p_{4} \right) \times
\hspace{1.6cm}\nonumber \\
\times e^{-\frac{\lambda^{4}}{4}\left|k_{s\mu\nu}k_{s}^{\;\mu\nu}\right|}\left\{
e^{-\frac{\lambda^{4}}{4}\left(p_{1\mu}p_{2\nu}+p_{3\mu}p_{4\nu}\right)^{2}}
\cosh\left[\frac{\lambda^{4}}{2}k_{s\mu\nu}\left(p_{1}^{\;\mu}p_{2}^{\;\;\nu}+p_{3}^{\;\;\mu}p_{4}^{\;\;\nu}\right)\right]
\right. \nonumber \\
\left.
+e^{-\frac{\lambda^{4}}{4}(p_{1\mu}p_{2\nu}-p_{3\mu}p_{4\nu})^{2}}
\cosh\left[\frac{\lambda^{4}}{2}k_{s\mu\nu}\left(p_{1}^{\;\mu}p_{2}^{\;\;\nu}-p_{3}^{\;\;\mu}p_{4}^{\;\;\nu}\right)\right]
\right. \nonumber \\
\left.
+e^{-\frac{\lambda^{4}}{4}(p_{1\mu}p_{3\nu}+p_{2\mu}p_{4\nu})^{2}}
\cosh\left[\frac{\lambda^{4}}{2}k_{s\mu\nu}\left(p_{1}^{\;\mu}p_{3}^{\;\;\nu}+p_{2}^{\;\;\mu}p_{4}^{\;\;\nu}\right)\right]
\right. \nonumber \\
\left.
+e^{-\frac{\lambda^{4}}{4}(p_{1\mu}p_{3\nu}-p_{2\mu}p_{4\nu})^{2}}
\cosh\left[\frac{\lambda^{4}}{2}k_{s\mu\nu}\left(p_{1}^{\;\mu}p_{3}^{\;\;\nu}-p_{2}^{\;\;\mu}p_{4}^{\;\;\nu}\right)\right]
\right. \nonumber \\
\left.
+e^{-\frac{\lambda^{4}}{4}(p_{1\mu}p_{4\nu}+p_{2\mu}p_{3\nu})^{2}}
\cosh\left[\frac{\lambda^{4}}{2}k_{s\mu\nu}\left(p_{1}^{\;\mu}p_{4}^{\;\;\nu}+p_{2}^{\;\;\mu}p_{3}^{\;\;\nu}\right)\right]
\right. \nonumber \\
\left.
+e^{-\frac{\lambda^{4}}{4}(p_{1\mu}p_{4\nu}-p_{2\mu}p_{3\nu})^{2}}
\cosh\left[\frac{\lambda^{4}}{2}k_{s\mu\nu}\left(p_{1}^{\;\mu}p_{4}^{\;\;\nu}-p_{2}^{\;\;\mu}p_{3}^{\;\;\nu}\right)\right]
\right\} \; ,
\end{eqnarray}
%
%
where we have defined
\begin{eqnarray}\label{kmunus}
k_{s\mu\nu}:=k_{1\mu\nu}+k_{2\mu\nu}
+k_{3\mu\nu}+k_{4\mu\nu} \; .
\end{eqnarray}
The expressions (\ref{Deltapk}) and (\ref{VertexV4}) are the Feynman rules of the
$\phi^{4}\star$ DFRA model. For radiative corrections of the perturbative series,
lines and vertex are represented in the momentum space by those expressions. Clearly,
the vertex expression shows that the external total momentum associated to extra dimension
$\theta$ is not conserved, that is, $k_{s\mu\nu} \neq 0$, while the total usual momentum $p^{\mu}$
is conserved.


\section{Conclusions}

The idea of noncommutativity brings hope to the elimination of divergences that plague QFT.  Snyder was the first one who published a way to deal with these ideas but Yang showed that the divergences were still there.  This result provoke an hibernation of Snyder work in particular and of the noncommutativity concepts in general for more than forty years.  Calculations concerning string theory algebra demonstrated that nature can be NC.  Since string theory is one of the candidates to unify QM with general relativity, noncommutativity concepts were vigorously reborn through a huge and dynamical literature.

To describe some NC aspects, one way that can be used is to analyze noncommutativity through the Moyal-Weyl product where the standard product of two or more fields is substituted by a star product.  In this case, the mathematical consistency of this star product is guaranteed because the NC parameter is constant (\cite{jhep} and references therein).  However, there are NC versions, not using the Moyal-Weyl product, where the NC parameter is not a constant.  We can also introduce noncommutativity through the so-called Bopp shift.  These ones are the most popular realizations of the NC concepts.

In this paper we work with an alternative and ingenuous formulation of NC theory developed recently, motivated by the ideas that in the early Universe, the spacetime may be NC.  This, by the way, is one of the main reasons that fueled NC research.  In this formulation (DFR) the NC parameter is a coordinate of spacetime.  So we have, in a D=4 Minkowski spacetime, for example, a NC space with six $\theta^{\mu\nu}$ coordinates, namely $D(D-1)/2$ NC coordinates.  In this NC spacetime, the $\theta^{\mu\nu}$-coordinate has its conjugated momenta $k_{\mu\nu}$ (DFRA), where we are interested in studying the consequences of the propagation in the $\theta$-direction. Besides, to work with a ten dimensional NC spacetime can disclose new physics beyond the Standard Model.

On the other hand, to work with NCQM we need only three space coordinates and consequently the NC sector has three coordinates also. These both sets combined are the so-called DFRA space which has been developed through these last few years using the standard concepts of QM and constructing an extension of the Hilbert space.

Here we showed that the alternative Feynman vision for QM can be treated in this DFRA space. We used this new formalism to quantize the NC free particle and the NC harmonic oscillator. After that we also constructed the DFRA action of a scalar field with a self-interaction $\phi^{4}\star$. From the action we have obtained the propagators and vertex $\phi^{4}\star$ in the momentum space. The detailed analysis of the radiative corrections of this model is the main motivation of the next paper.

As a final remark we would like to say that the formalism developed here is very unusual concerning the fact that it was totally constructed within a NC space. We did not introduce (by hand) any NC parameter such that its elimination recover the commutative theory. To recover its commutative behavior, we have to perform a dimensional reduction in this $(x+\theta)$-space which has nine dimensions.  Since some thermodynamics were considered and some objects like transition amplitude and partition function with NC coordinates were calculated, we believe that the construction of a NC thermofield dynamics can be the next move in this direction.  Some cosmological models like black holes and wormholes can be a target for this analysis. The idea would be, instead of introducing noncommutativity in such models existing in commutative spacetime, to construct black holes and wormholes in a NC spacetime like the one discussed here.


\begin{thebibliography}{30}

\bibitem{hawking}   S. W. Hawking, Nucl. Phys. B 144 (1978) 349, and references therein.

\bibitem{snyder47} H. S. Snyder, {\it Phys. Rev.} {\bf 71} (1947)  38.

\bibitem{yang47}  C. N. Yang, {\it Phys. Rev.} {\bf 72} (1947) 874.

\bibitem{seibergwitten99} N. Seiberg and E. Witten, {\it JHEP} {\bf 9909} (1999) 032.

\bibitem{QG}  L. Alvarez-Gaum\'{e}, F. Meyer and M. A. Vazquez-Mozo, {\it Nucl. Phys.}
B {\bf 753} (2006) 92;
\\
X. Calmet and A. Kobakhidze, {\it Phys. Rev.} D {\bf 72} (2005) 045010;
\\
E. Harikumar and V. Rivelles, {\it Class. Quantum Gravity} {\bf 23} (2006) 7551;
\\
M. R. Douglas and N. A. Nekrasov, {\it Rev. Mod. Phys.} {\bf 73} (2001) 977;
\\
R. J. Szabo, Class. Quant. Grav. {\bf 23} (2006) R199;
\\
R. J. Szabo, {\it Quantum Gravity, Field Theory and Signatures of NC Spacetime}, arXiv:0906.2913;
\\
F. M\"uller-Hoissen, {\it NC geometries and gravity} {\it AIP Conf. Proc.} {\bf 977} (2008);
\\
V. Rivelles, {\it Phys. Lett.} B {\bf 558} (2003) 191;
\\
H. Steinacker, {\it JHEP}
{\bf 0712} (2007) 049, {\it Nucl. Phys. B} {\bf 810} (2009) 1;
\\
R. Banerjee, H. S. Yang, {\it Nucl. Phys.} B {\bf 708} (2005) 434;
\\
M.R.Douglas and C. Hull, {\it JHEP} {\bf 9802} (1998) 008;
\\
P. A. Horvathy and M. S. Plyushchay, {\it JHEP} {\bf 0206} (2002) 033; {\it Phys. Lett.} B {\bf 595} (2004) 547; {\it Nucl. Phys.} B {\bf 714} (2005) 269;
\\
M. S. Plyushchay, {\it JHEP} {\bf 0903} (2009) 034;
\\
R. Banerjee, B. Chakraborty and K. Kumar, {\it Phys. Rev.} D {\bf 70} (2004) 125004;
\\
A. Kempf, G. Mangano and R. B. Mann, {\it Phys. Rev.} D {\bf 52} (1995) 1108.

\bibitem{Szabo03} R. Szabo, {\it Phys. Rep.} {\bf 378} (2003) 207.

\bibitem{DFR} S. Doplicher, K. Fredenhagen and J. E. Roberts, {\it Phys. Lett.} B {\bf 331}
(1994) 29; {\it Commun. Math. Phys.} {\bf 172} (1995) 187.

\bibitem{Morita} H. Kase, K. Morita, Y. Okumura and E. Umezawa, {\it Prog. Theor. Phys.}
 {\bf 109} (2003) 663; K. Imai, K. Morita and Y. Okumura, {\it Prog. Theor. Phys.}
{\bf 110} (2003) 989.

\bibitem{Amorim1} R. Amorim, {\it Phys. Rev.  Lett.} {\bf 101} (2008) 081602.

\bibitem{amo} E. M. C. Abreu, A. C. R. Mendes, W. Oliveira and A. Zagirolamim, {\it SIGMA} {\bf 6} (2010) 083.

\bibitem{EMCAbreuMJNeves2011} E. M. C. Abreu and M. J. Neves, {\it Int. J. Mod. Phys.} A {\bf 27} (2012) 1250109.

\bibitem{Carlson} C. E. Carlson, C.D. Carone and N. Zobin, {\it Phys. Rev.} D {\bf 66}
(2002) 075001.

\bibitem{Carone} C. D. Carone and H. J. Kwee, {\it Phys. Rev.} D {\bf 73} (2006) 096005.

\bibitem{Conroy2003} J. M. Conroy, H. J. Kwee and V. Nazarayan, {\it Phys. Rev.} D {\bf 70}, 034017 (2004).

\bibitem{Saxell} S. Saxell, {\it Phys. Lett.} B {\bf 666} (2008) 486.

\bibitem{Amorim4}  R. Amorim, {\it Phys. Rev.} D {\bf 78} (2008) 105003.

\bibitem{Amorim5}  R. Amorim, {\it J. Math. Phys.} {\bf 50} (2009) 022303.

\bibitem{Amorim2}  R. Amorim,  {\it J. Math. Phys.} {\bf 50}  (2009) 052103.

\bibitem{Iorio}  A. Iorio and T. Sykora, {\it Int. J. Mod. Phys.} A {\bf 17} (2002) 2369;
\\
A. Iorio, {\it Phys. Rev.} D {\bf 77} (2008) 048701.


\bibitem{1}   F. G. Scholtz, L. Gouba, A. Hafver and C. M. Rohwer, {\it J. Phys.} A {\bf 42} (2009) 175303.

\bibitem{2}   C. M. Rohwer, K. G. Zloshchastiev, L. Gouba and F. G. Scholtz, {\it J. Phys.} A {\bf 43} (2010) 345302.

\bibitem{3}   A. Samilagic and E. Spallucci, {\it J. Phys.} A {\bf 36} (2003) L467.

\bibitem{4}   S. Gangopadhyay and F. G. Scholtz, {\it Phys. Rev. Lett.} {\bf 102} (2009) 241602.

\bibitem{15a}   S. Gangopadhyay and F. G. Scholtz, {\it Free particles on noncommutative plane - a coherent state path integral}, arXiv: 0812.3474.


\bibitem{5}   G. Mangano, {\it J. Math. Phys.} {\bf 39} (1998) 2584.

\bibitem{6}   M Hale, {\it J. Geom. Phys.} {\bf 44} (2002) 115.

\bibitem{7}   C. Acatrinei, {\it JHEP} {\bf 0109} (2001) 007.

\bibitem{8}   A. Jahan, {\it Braz. J. Phys.} {\bf 38} (2008) 144.

\bibitem{9}   A. Jahan, Fizika B 18 (2009) 4, 189.

\bibitem{10}   I. Chepelev and C. Ciocarlia, {\it JHEP} {\bf 0306} (2003) 031.

\bibitem{11}   K. Okuyama, {\it JHEP} {\bf 0003} (2000) 016.

\bibitem{12}   K. Fujikawa, {\it Phys. Rev.} D {\bf 70} (2004) 085006.

\bibitem{13}   B. Dragovich and Z. Raki{\'{c}}, {\it Theor. Math. Phys.} {\bf 140} (2004) 1299.

\bibitem{14}   B. Dragovich and Z. Raki{\'{c}}, {\it Noncommutative quantum mechanics with path integral}
arXiv: hep-th/0501231; {\it Path integral approach to noncommutative quantum mechanics}, arXiv: hep-th/0401198.

\bibitem{15}   D. Gitman and V. G. Kupriyanov, {\it Eur. Phys. J.} C {\bf 54} (2008) 325.

\bibitem{28}   F. Khelili, {\it Path Integral Quantization of Noncommutative Complex Scalar Field}, arXiv: 1109.4741.

\bibitem{Chaichian}  M. Chaichian, M. M. Sheikh-Jabbari and A. Tureanu, {\it Phys. Rev. Lett.} {\bf 86} (2001), 2716-2719.

\bibitem{Gamboa}  J. Gamboa, M. Loewe and J. C. Rojas, {\it Phys. Rev.} D {\bf 64} (2001), 067901.

\bibitem{Kokado}   A. Kokado, T. Okamura and T. Saito, {\it Phys.  Rev.} D {\bf 69} (2004), 125007.

\bibitem{Kijanka}  A. Kijanka and P. Kosinski, {\it Phys. Rev.} D {\bf 70} (2004), 127702.

\bibitem{Calmet} X. Calmet, {\it Phys. Rev.} D {\bf 71} (2005), 085012;
\\
X. Calmet and M. Selvaggi, {\it Phys. Rev.} D {\bf 74} (2006), 037901.

\bibitem{jhep}  E. M. C. Abreu, M. V. Marcial, A. C. R. Mendes and W. Oliveira and G. Oliveira-Neto {\it JHEP} {\bf 1205} (2012) 144.












\end{thebibliography}
\end{document}